\documentstyle[aps, prb,twocolumn, floats, epsfig]{revtex}

\begin{document}

\newcommand{\gsim}{
\,\raisebox{0.35ex}{$>$}
\hspace{-1.7ex}\raisebox{-0.65ex}{$\sim$}\,
}

\newcommand{\lsim}{
\,\raisebox{0.35ex}{$<$}
\hspace{-1.7ex}\raisebox{-0.65ex}{$\sim$}\,
}

\newcommand{\const}{ {\rm const} }
\newcommand{\arctanh}{ {\rm arctanh} }

\bibliographystyle{prsty}

\title{  
\begin{center}
\small PHYSICAL REVIEW B {\normalsize 66}, 174419 (2002)
\end{center}
Hysteretic properties of a magnetic particle with strong surface anisotropy
}      
\author{
H Kachkachi\cite{e-kac} and M Dimian 
}

\address{
Laboratoire de Magn\'{e}tisme et d'Optique, Universit\'e de Versailles St.
Quentin, \\
45 av. des Etats-Unis, 78035 Versailles, France\\
\smallskip
\today
\bigskip\\
\parbox{14.2cm}
{\rm
We study the influence of surface anisotropy on the zero-temperature
hysteretic properties of a small single-domain 
ferromagnetic particle, and investigate limiting cases where
deviations from the Stoner-Wohlfarth model are 
observed due to non-uniform reversal of the particle's magnetization.
We consider a spherical particle with simple cubic
crystal structure, a uniaxial anisotropy for core spins and
radial anisotropy on the surface.
The hysteresis loop is obtained by solving the local (coupled)
Landau-Lifshitz equations for classical spin vectors.
We find that when the surface anisotropy constant $K_s$ assumes large
values, e.g. of the order of
the exchange coupling $J$, large deviations are observed with
respect to the Stoner-Wohlfarth model in the hysteresis loop and thereby
the limit-of-metastability curve, since in this case the magnetization
reverses its direction in a non-uniform manner via a progressive switching of
spin clusters.  
This characteristic value of $K_s$ depends on
the surface-to-volume ratio of exchange coupling and the angle between
the applied field and core easy axis.
\smallskip
\begin{flushleft}
PACS number(s): 75.50.Tt - 75.30.Pd - 75.10.Hk
\end{flushleft}
} 
} 
\maketitle

\section{Introduction}
Surface effects have a strong bearing on the properties of small magnetic
systems, and entail large deviations from the bulk behavior. It was shown
in \cite{Kachkachi} that the magnetic disorder on
the surface caused by surface anisotropy is long ranged, which implies that
even the spins in the core of a very small magnetic particle (2 nm) render a
magnetization that deviates from the bulk value. It will be useful to
understand surface effects in magnetic materials in order to control
their properties which are relevant for technological applications. One such
property is the coercive field as it gives indications on the relaxation
time of the magnetization and thereby on the stability of the information
stored on magnetic media.

Surface effects are due to the breaking of crystal-field symmetry, and this
is a local effect. So, in order to study such effects one has to resort to
microscopic theories, unlike the macroscopic Stoner-Wohlfarth (SW)
model \cite{SW}, which are capable of distinguishing between different
atomic environments and taking account of physical parameters such as bulk
and surface anisotropy, exchange and dipole-dipole interactions.
Unfortunately, this leads to difficult many-body problems which can only be
dealt with using numerical approaches. 

This work deals with the effect of strong
surface anisotropy on the hysteretic properties (hysteresis loop and
limit-of-metastability curve, the so-called SW astroid), of a
single-domain spherical
particle (with free surfaces), a simple cubic (sc) crystal
structure, a uniaxial anisotropy in the core, and radial single-site
anisotropy for spins on the boundary.
The hysteresis loop and thereby the critical field are computed by
solving, at zero temperature, the  
local Landau-Lifshitz equations derived from the classical anisotropic
Dirac-Heisenberg model in field, subjected to a local condition (see below)
accounting for the minimization of energy with respect to local
rotations of each spin in the particle.
In Ref.\cite{DimitrovWysin} the same method was used for studying the
hysteretic properties of models of nanoparticles, where the anisotropy
was either random in the whole particle or taken only on the surface,
and the analysis was restricted to the hysteresis loop. 

In this paper, we use an improved version of the method mentioned above
including a global-rotation condition on the resultant magnetic moment
of the particle in addition to the local condition (see
\cite{HK+DG2}). We compute the  
hysteresis loop and infer from it the limit-of-metastability curve (SW astroid),
and compare with the SW model especially when the surface anisotropy constant
assumes large values, e.g. $K_s/J \sim 1$. This study has allowed us to
investigate the limit of validity of the SW model for very small
magnetic particles where surface anisotropy plays a determinant
role, and whose magnetization no longer switches in a coherent way. 

Our method, based on the numerical solution of the
Landau-Lifshitz equation at zero temperature, is checked against the
SW semi-analytical results in two 
limiting cases of the exchange coupling with different distributions
of anisotropy axes. We first consider a
single-domain particle with a macroscopic magnetic moment resulting
from very strong exchange interaction. This is equivalent to the
SW one-spin problem with uniaxial anisotropy. 
A second test deals with the case 
of a square particle of non-interacting spins all with randomly distributed
easy axes. This model mimics an assembly of mono-dispersed single-domain
nanoparticles with a random distribution of their easy axes embedded
in a 2d non-magnetic matrix.

The plan of this work is as follows: we first define our model (Hamiltonian and
physical parameters), present the method used for 
computing the hysteresis loop, and test it against the semi-analytical
results of SW model. Then, we discuss our results for a spherical
particle in terms of exchange coupling, particle's size, and
surface anisotropy by varying, in turn, one of them while
keeping the other two fixed. We also study the situation with (intra)
surface exchange coupling different from that in the core of the
particle. A short account of the present work can be found in
Ref.~\cite{MD+HK}. 
It is worth mentioning though that in fact only anisotropy and
exchange coupling on the surface can be considered as free parameters
as there are so far no definite experimental estimations thereof. 

\section{Model Hamiltonian}

We consider the following classical anisotropic Dirac-Heisenberg model

\begin{equation}
{\cal H}=-\sum\limits_{\left\langle i,j\right\rangle }J_{ij}{\bf S}_{i}{\bf \cdot
S}_{j}-(g\mu _{B}){\bf H\cdot }\sum\limits_{i=1}^{{\cal N}}{\bf S}%
_{i} + H_{an}, \label{DH}
\end{equation}
where ${\bf S}_{i}$ is the unit spin vector on site $i,$ ${\bf H}$ is the
uniform magnetic field applied in a direction $\psi$ with
respect to the reference $z$ axis, ${\cal N}$ is the
total number of spins (core and surface), and in the sequel $D$ will denote the
particle's diameter. $J_{ij} (= J>0)$  
is the strength of the nearest-neighbor exchange interaction, which will be taken in
our calculations the same everywhere inside the particle, unless
otherwise specified (see Fig.\ \ref{hcvsks2} et seq); $H_{an}$ is the
uniaxial anisotropy energy
\begin{equation}\label{uaa}
H_{an} = -\sum\limits_{i}K_{i}({\bf S}_{i}{\bf \cdot e}_{i})^{2},  \label{Ham}
\end{equation}
with easy axis ${\bf e}_{i}$ and constant $K_{i}>0$.
This anisotropy term contains either of the two
contributions stemming from the core and surface, and depends on the system
under consideration. For instance, for the 
2d model (which serves as a test of our calculations by comparison with
the SW model) all spins (core and surface) have the same
anisotropy constant but randomly distributed axes.
In the case of a spherical particle, all core spins are attributed
the same constant $K_c$ and all surface spins are attributed the constant
$K_s$. 
Moreover, core spins will have an easy axis along the $z$ axis, whereas
a surface spin is assumed to have its anisotropy axis along the 
radial direction, see \cite{TSA} and many references therein.

A more physically appealing microscopic model of surface anisotropy was
provided by N\'eel\cite{Neel},
\begin{equation}\label{NSA}
H_{an}^{Neel} = -K_s\sum\limits_{i}\sum\limits_{j=1}^{z_i}({\bf S}_{i}{\bf \cdot
e}_{ij})^{2},
\end{equation}
where $z_i$ is the coordination number of site $i$ and ${\bf e}_{ij}={\bf
r}_{ij}/r_{ij}$ is the unit vector connecting the site $i$ to its nearest
neighbors. This model is more realistic since the anisotropy at a given
site occurs only when the latter looses some of its neighbors, i.e. when it
is located on the boundary. 
However, the extra sum on nearest-neighbors in (\ref{NSA}) makes this model
less practical for numerical calculations, especially those that are time
consuming, such as the SW astroid. 
So in this paper we restrict ourselves to the model of radial single-site
anisotropy on the surface.
In Ref.~\cite{HKDG_NSA}, we have developed an analytical 
theory, together with the numerical method used here, for weak surface
anisotropy and studied this model and compared it with the radial-anisotropy
model. 

A remark is in order concerning the dipole-dipole interactions inside
the particle. It is well known \cite{Akhiezer} that these
relativistic interactions lead to two 
contributions, a first term that is an integral over the volume of the
particle, and a second one over the surface. The latter represents the
magnetostatic energy. However, it has been shown \cite{Hahn} that in very small
particles the first contribution is negligible as compared with the
contribution of exchange interactions. On the other hand, the second
contribution plays the role of shape anisotropy, which for a spherical
particle yields an irrelevant constant. 
Therefore, in our case of very small spherical particles, where the
effect of surface anisotropy constant is most important, which is one of
the main issues of the present work, the volume term is negligible and the 
shape anisotropy is absent.
\section{Method of calculation of the hysteresis loop}

Different models of a nanoparticle are studied. In each case, we simulate
the lattice with sc crystal structure, and then assign to each
site a length-fixed three-component spin vector. For the
calculation of the hysteresis loop we start with a magnetic configuration
where all spins are pointing in the same direction $-z$, which
corresponds to the saturation state. 
The hysteresis loop is due to the existence of metastable states in the system.
Starting from the initial configuration and applied field, the integration
of the Landau-Lifshitz equation (see below) tends towards a new
configuration that is an energy minimum.

Let us now establish the Landau-Lifshitz equations for the magnetic moments. We
choose $K_{c}$ as the energy scale and normalize the other physical
constants accordingly, i.e., 
\begin{equation}
t\rightarrow \frac{2K_{c}}{\hbar}\times t,\quad {\bf h\equiv }\frac{(g\mu
_{B})}{2K_{c}}\times {\bf H}.  \label{3}
\end{equation}
Then, the Landau-Lifshitz (LL) equation for a spin ${\bf S}_{i}$ at site $i,$
reads 
\begin{equation}
\frac{d{\bf S}_{i}}{dt}=-{\bf S}_{i}\times {\bf h}_{i}^{eff}-\alpha {\bf S}%
_{i}\times \left( {\bf S}_{i}\times {\bf h}_{i}^{eff}\right)  \label{LLE}
\end{equation}
where $\alpha (\sim 1)$ is the damping parameter and ${\bf h}_{i}^{eff}$ is the
effective field acting on the spin ${\bf S}_{i}$ and is given by 
\begin{equation}
{\bf h}_{i}^{eff}={\bf h+}\frac{1}{2K_{c}}\sum\limits_{j=1}^{z_i}J_{ij}{\bf S}_{j}+%
{\bf h}_i^{an}  \label{EffField}
\end{equation}
where ${\bf h}_{i}^{an}\equiv -(\partial H_{an}/\partial {\bf S}_{i})/2K_c$, with $%
H_{an}$ given in Eq.\ (\ref{uaa}), $z_i$ is the coordination number of
site $i$. In the sequel, we will use the reduced
parameters, $j\equiv J/K_{c},$ $k_{s}\equiv K_{s}/K_{c}.$ Therefore, for
each site $i$ we arrive at three coupled equations (for $%
S_{i}^{x},S_{i}^{y},S_{i}^{z}$), and because of the second term in (\ref
{EffField}) we actually obtain a system of $3{\cal N}$ (local) coupled
equations. 
We emphasize that it is more convenient to use spherical
coordinates (for each spin) instead of the Cartesian ones. Indeed, owing to
the fact that the spins are of constant length, this reduces the number
of individual (for each spin) equations to two instead of three, 
\begin{eqnarray}
&&\dot{\theta}_{i}=(h_{\varphi }^{eff}+\alpha h_{\theta }^{eff})_{i}
\label{LLEsc} \\
&&\dot{\varphi}_{i}=\left( -h_{\theta }^{eff}+\alpha h_{\varphi
}^{eff}\right) _{i}/\sin \theta _{i},  \nonumber
\end{eqnarray}
where $h_{\theta }^{eff}\equiv -\partial {\cal H}/\partial \theta
,h_{\varphi }^{eff}\equiv -\partial {\cal H}/\partial \varphi$, are the
polar components of the effective field. For the one-spin problem, these are
obtained by direct differentiation of the energy written in spherical
coordinates, whereas for a particle it is not possible to obtain a tractable analytical
expression of the energy in spherical coordinates, so $h_{\theta }^{eff}$
and $h_{\varphi }^{eff}$ are written in terms of the time derivatives of the
Cartesian components of ${\bf h}_{i}^{eff}$ in (\ref{EffField}). Using Eq.\ (%
\ref{LLEsc}) instead of (\ref{LLE}) allows for a gain of computer time, but
this method encounters stability problems specific to the spherical
coordinates, because of the factor $1/\sin \theta $ in (\ref{LLEsc}), which
diverges as $\theta \rightarrow 0,\pi $, and hence a special care is
required when numerically handling these equations.

After having constructed the magnetic structure (lattice and spin vectors on
it), we apply a magnetic field ${\bf H}$ at some angle $\psi$ with
respect to the reference $z$ axis, with
values chosen in a regular mesh. Then we calculate the local 
effective field for all spins and thereby the right-hand sides of the LL
equations (\ref{LLEsc}) and proceed with the time integration. As this is
done, the total energy in Eq.(\ref{DH}) smoothly decreases, and some
criterion must be used for stopping the integration for the given value of
the applied field and moving to the next value. In our calculations we
proceed to the next field value when 
\begin{equation}
\frac{1}{\cal N}\sum\limits_{i=1}^{{\cal N}}\left| \frac{d{\bf S}_{i}}{dt}\right|
<\varepsilon ,  \label{condition1}
\end{equation}
which implies that the system is close to a stationary state, $\varepsilon $
being a small parameter of the order of $10^{-5}-10^{-7}$. However, it was
shown in \cite{HK+DG2}, that this local condition, which accounts for the
minimization of energy with respect to local rotations (or small deviations)
of each spin, must be supplemented by a global condition on the
resultant magnetic moment so as to account for the global rotation of the
particle's magnetic moment. Obviously, for a single spin these two
conditions boil down to one and the same condition
(\ref{condition1}).

Next, the stationary state thus obtained is used as the initial
state for the next value of the field. Iteration of this process over a
sequence of applied fields, of given magnitude and direction $\psi$, renders
the hysteresis loop. For each value of this angle 
we determine the critical or switching field (see discussion
below). The whole procedure finally renders the critical or switching
field as a function of the angle $\psi$, which in
the case of critical field is the SW astroid.

\begin{figure*}[t]
\unitlength1cm 
\begin{picture}(9,10)(0,-10)
\centerline{\psfig{file=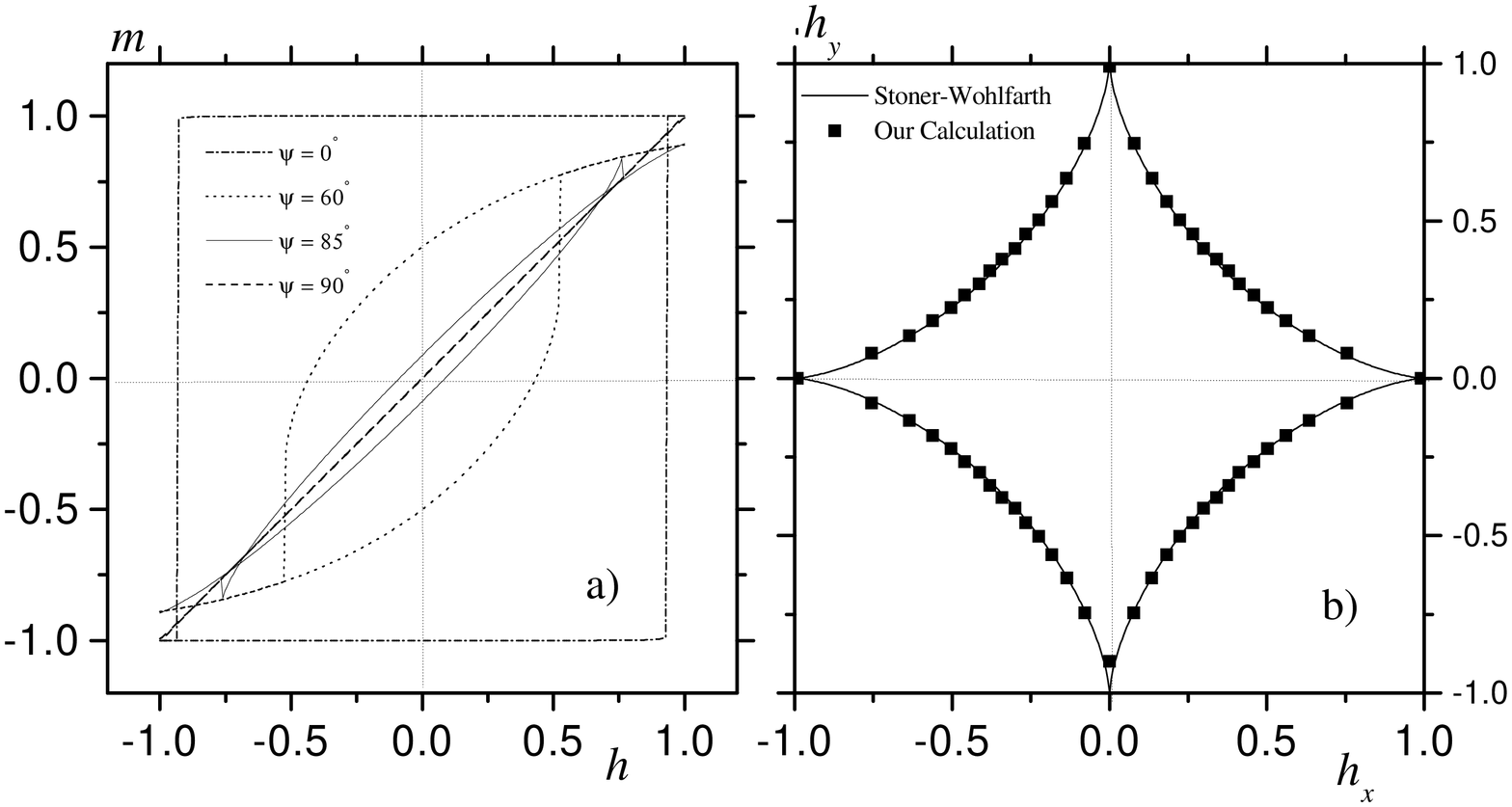, angle=-90, width=14cm}}
\end{picture}
\vspace{-2.0cm}
\caption{\label{swt1}
Left: (numerical) hysteresis loops for different values of $\psi $ increasing
inwards: $\psi =0,60^{\circ},85^{\circ},90^{\circ}$, for a $3^{3}$
particle with uniaxial anisotropy. For the sake of clarity the SW analytical
hysteresis loops have been omitted, since they exactly coincide with the computed
ones. Right: (numerical in squares and
analytical in full line) SW astroid for 
the same particle; $j=10$.
}
\end{figure*}

\begin{figure*}[!]
\unitlength1cm 
\begin{picture}(9,8)(0,-9)
\centerline{\psfig{file=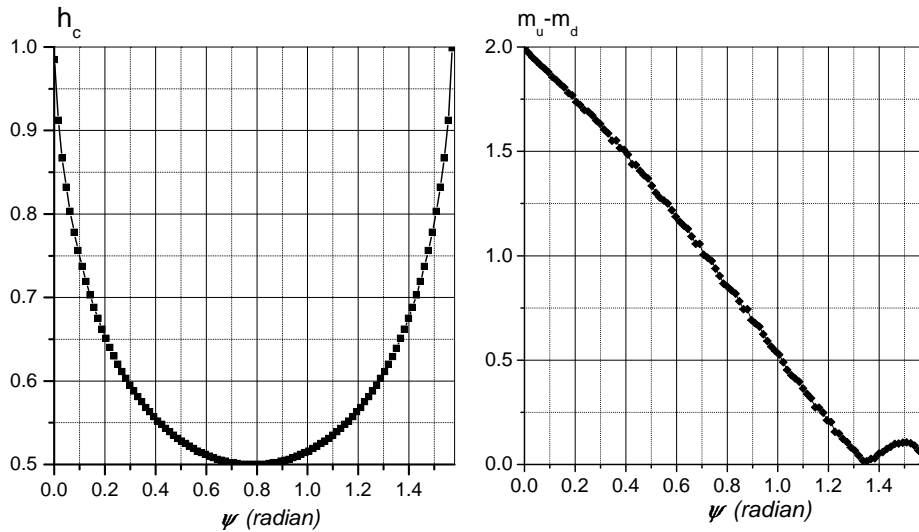, angle=-90, width=13cm}}
\end{picture}
\vspace{-1.5cm}
\caption{\label{hc_width}
One-spin problem. Left: critical field as function of $\psi$. 
Right: height of magnetization jump as function of $\psi$.}
\end{figure*}

As a test of this method, we considered a box-shaped 
particle with\cite{footnote3} ${\cal N}=3^{3}$, a sc structure, uniaxial 
anisotropy, and strong exchange interaction between spins inside the
particle, and computed the hysteresis loop for different values of the angle 
$\psi$ between the applied field and the easy axis. 
The results are shown in Fig.\ \ref{swt1} (left). Next, we present
in Fig.\ \ref{swt1} (right) the SW astroid, which separates the region
with two minima of energy from that with only one minimum.
We see that the SW results are exactly
reproduced by our calculations.
We have also computed the hysteresis loop of a square
particle of non-interacting spins $(J = 0)$ all with randomly
distributed easy axes. This is equivalent to an assembly of
mono-dispersed single-domain non-interacting particles with randomly
distributed easy axes in two dimensions. 
As expected, we find that the remanent magnetization is equal to 
$1/2$. 
%
%
For later reference, we plot in Figs.\ \ref{hc_width} the
critical field $h_{c}$ and the height of the magnetization jump (i.e. 
$m_{u}-m_{d}$), as functions of the angle $\psi $ between the direction of the
field and core easy axis. Obviously, $h_c(\psi)$ in Fig.\
\ref{hc_width} (left) is a well known result of the Stoner-Wohlfarth
model. 
%
%
On the other hand, we note that the height of magnetization jump has
an almost linear dependence
on $\psi$, except for the final portion $76^\circ < \psi < 90^\circ$,
which corresponds to cycles with crossing branches as exhibited by
the hysteresis for $\psi =85^\circ$ in Fig.\ \ref{swt1} (left), see
\cite{Hagedorn} for a discussion of this issue.

\section{Spherical particles: results and discussion}

\begin{figure}[t!]
\unitlength1cm
\begin{picture}(15,10)(-1,-9)
\centerline{\psfig{file=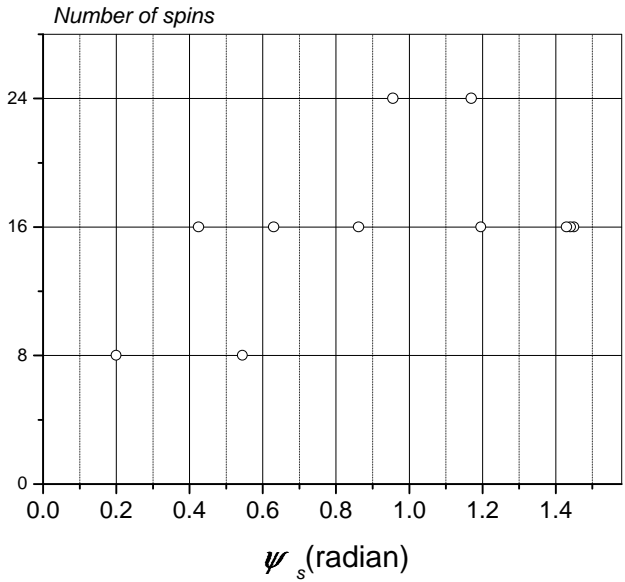, angle=-90, width=12cm}}
\end{picture}
\caption{\label{number}
Distribution of surface anisotropy axes versus the azimuthal angle $\psi_s$ for a
spherical particle with $D=10$ (${\cal N}=360$: 176 surface spins and 184 core spins).
}
\end{figure}

Here we consider a single-domain spherical particle of simple cubic (sc) structure with
uniaxial anisotropy in the core and anisotropy constant $K_{c}$, and radial
anisotropy on the surface with constant $K_{s}$. 
Our main goal here 
is to investigate the influence of 
surface anisotropy, both in direction and strength, on the hysteresis loop
and SW astroid. However, we will also study the
effect of exchange coupling and particle's size.
Again for later reference, we plot in Fig.\ \ref{number} the
distribution of surface anisotropy axes of the spherical particle as a
function of the azimuthal angle $\psi_s$ between a surface spin easy
axis and applied field.
%
%
\subsection{Effect of the exchange coupling $j$}
\begin{figure}[floatfix]
\unitlength1cm
\begin{picture}(9,8)(-1.25,-9)
\centerline{\psfig{file=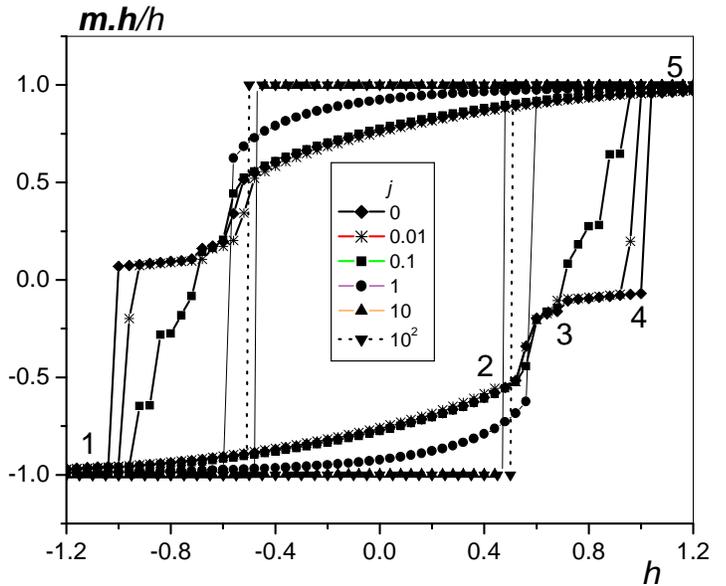, angle=-90, width=13cm}}
\end{picture}
\caption{\label{hystk1n10}
Hysteresis loop, i.e. plot of the magnetization projection on the field
direction as a function of the (reduced) field $h$, for $\psi =0$, $k_s=1$
and different values of $j$. ${\cal N}=360$. 
}
\end{figure}

\begin{figure*}[!]
\unitlength1cm 
\begin{picture}(4,4.5)(5.5,0)
\centerline{\psfig{file=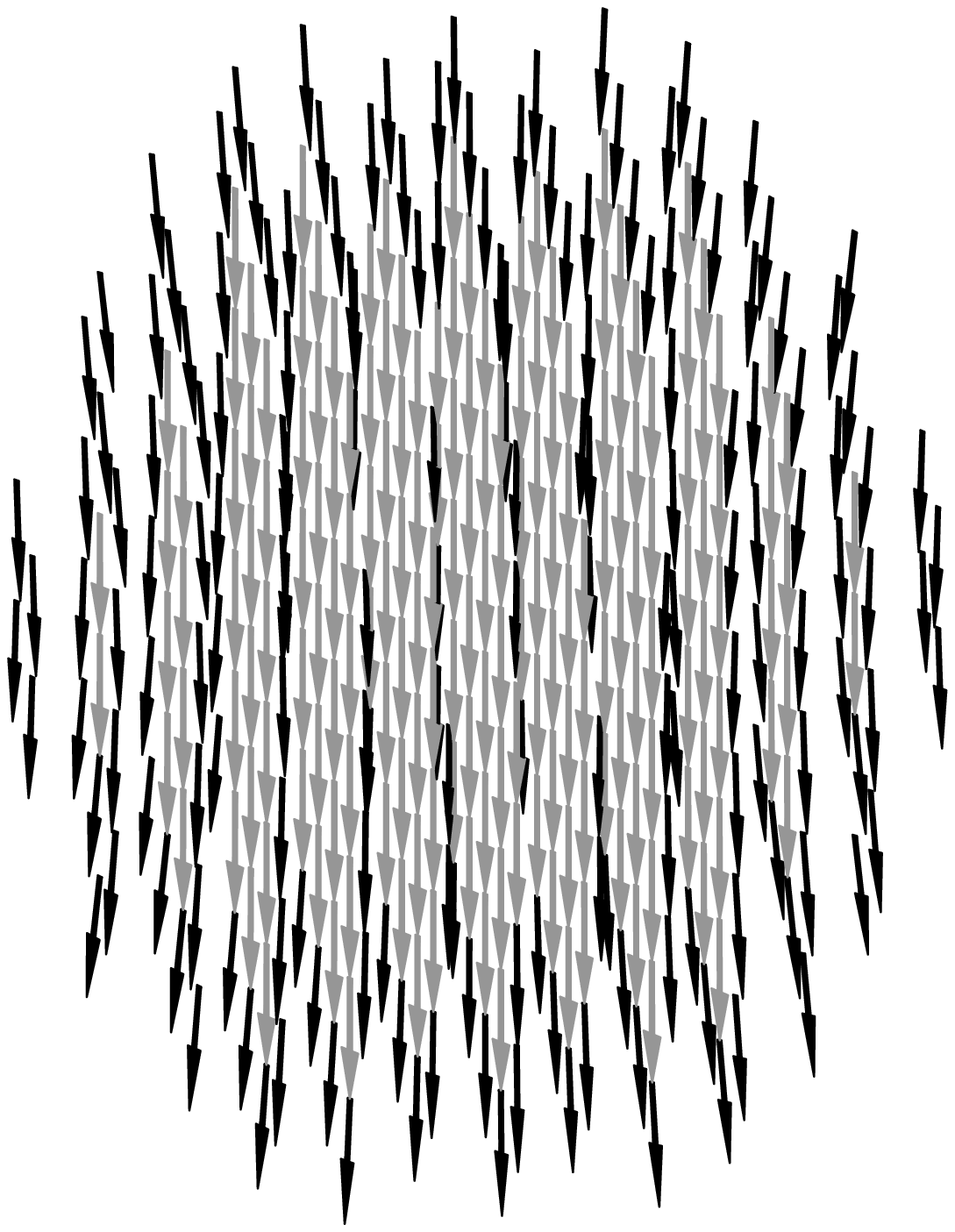, width=4.5cm, height=4.5cm}}
\end{picture}
\begin{picture}(4,4.5)(4.3,0)
\centerline{\psfig{file=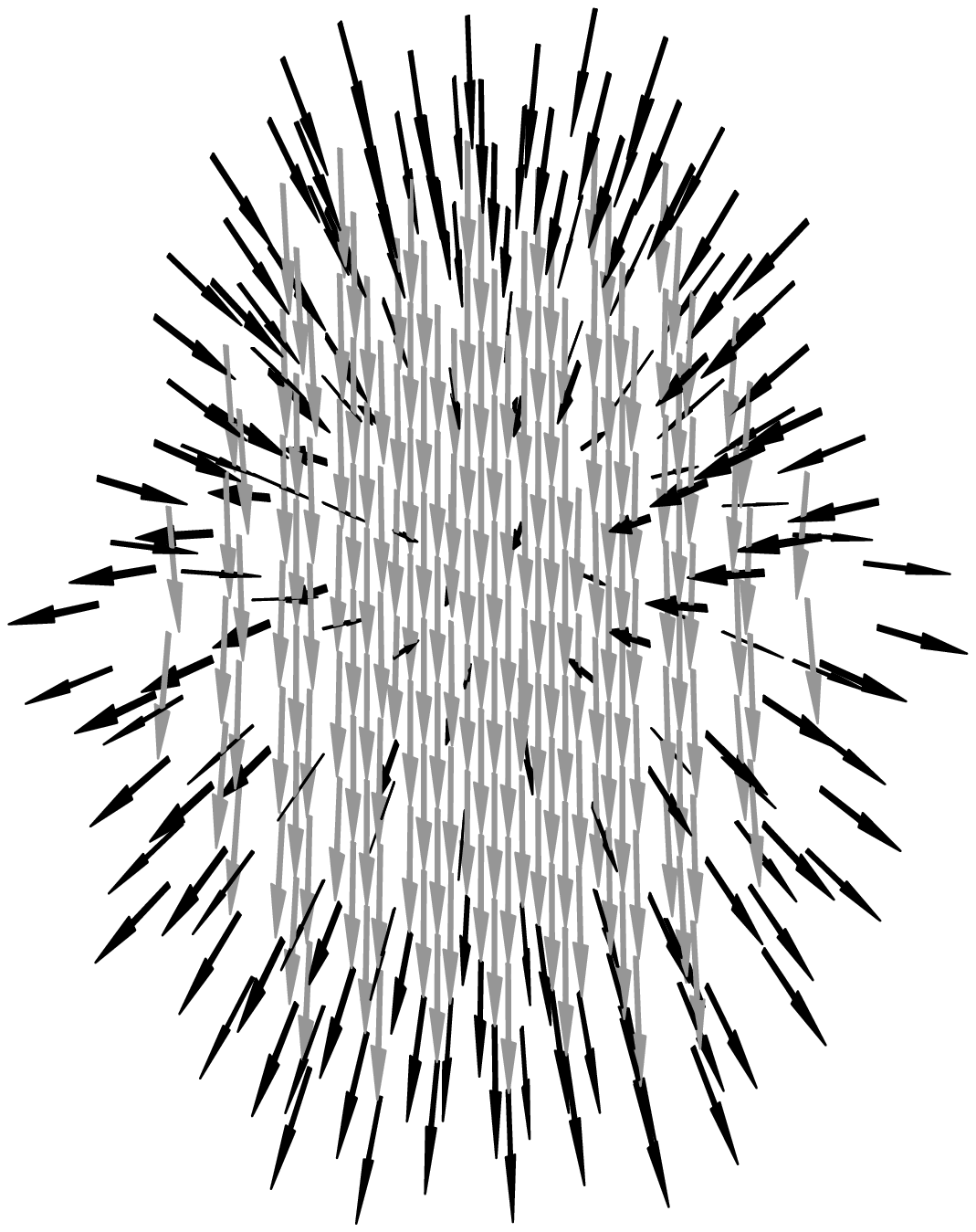, width=4.5cm, height=4.5cm}}
\end{picture}
\begin{picture}(4,4.5)(3.5,0)
\centerline{\psfig{file=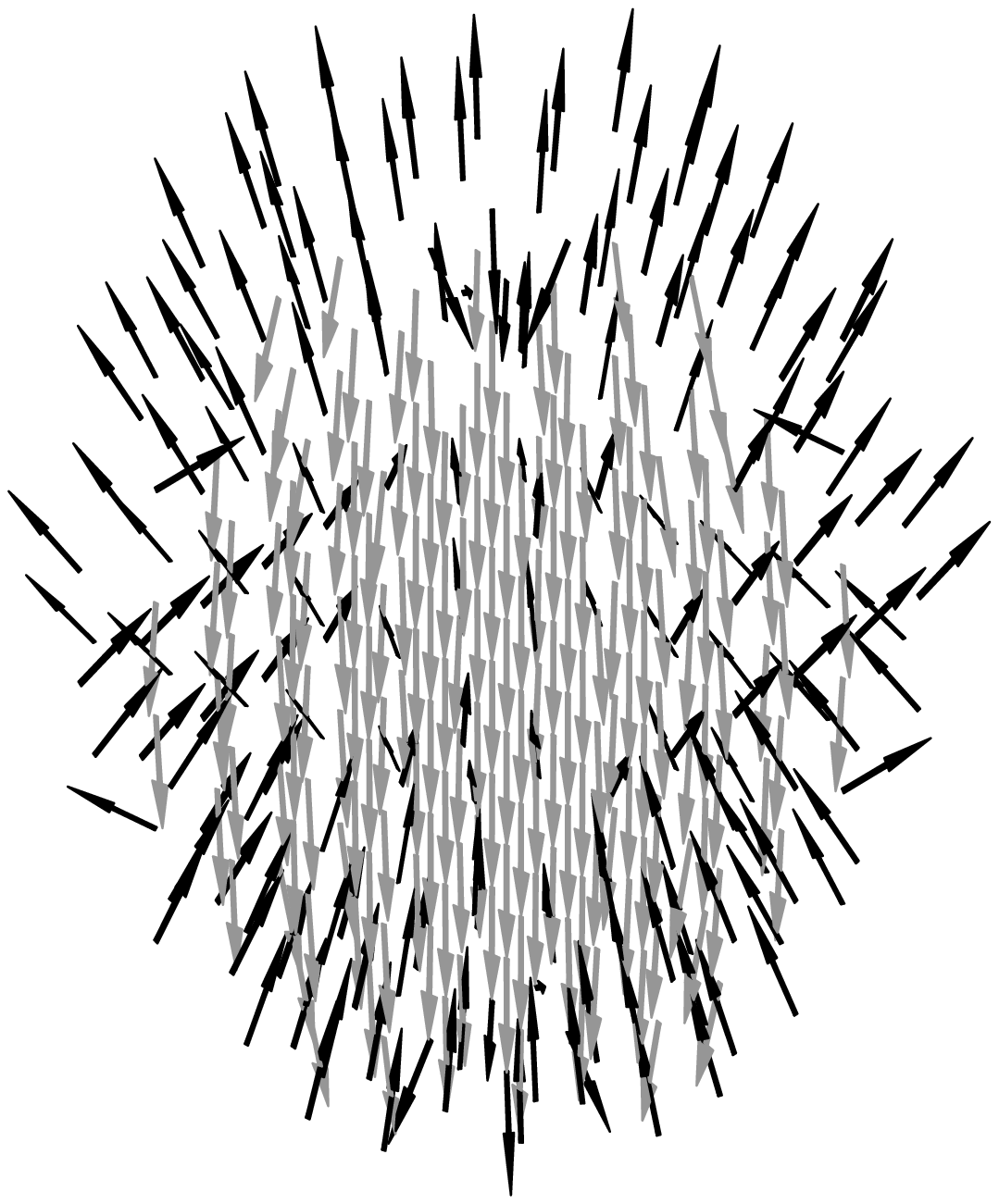, width=4.5cm, height=4.5cm}}
\end{picture}\\
\begin{picture}(4,4.5)(5.0,0)
\centerline{\psfig{file=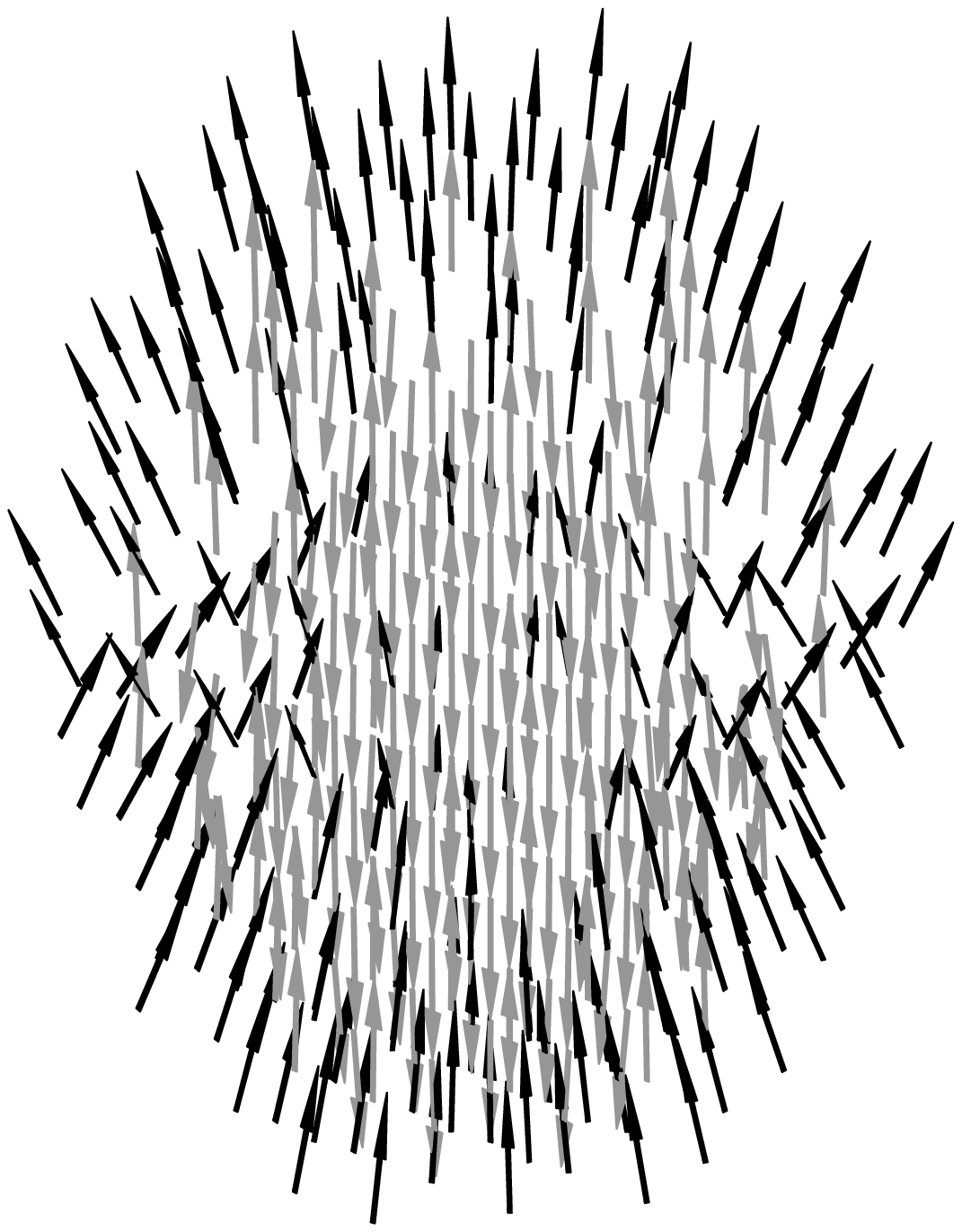, width=4.5cm, height=4.5cm}}
\end{picture}
\begin{picture}(4,4.5)(4,0)
\centerline{\psfig{file=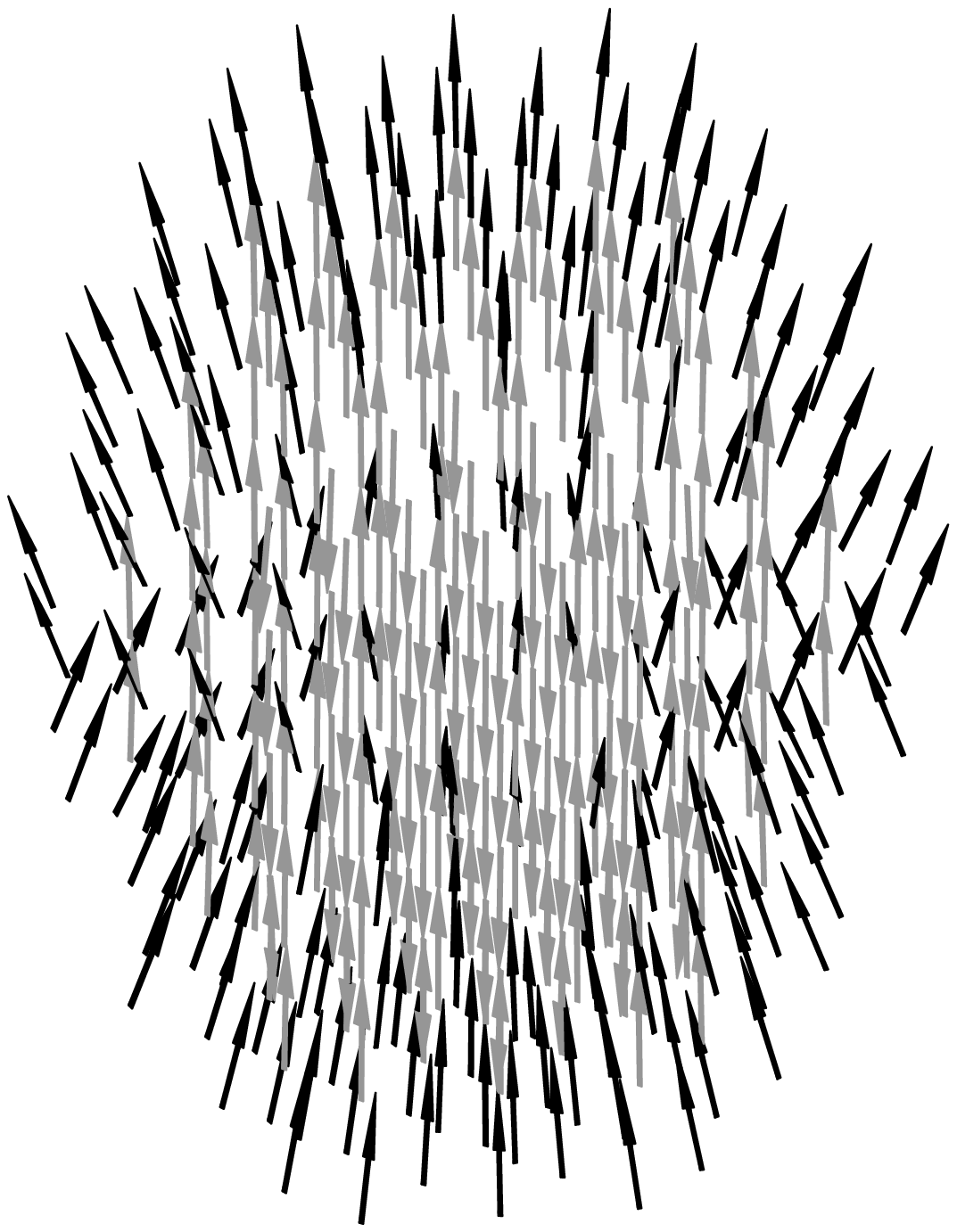, width=4.5cm, height=4.5cm}}
\end{picture}
\begin{picture}(4,4.5)(3.0,0)
\centerline{\psfig{file=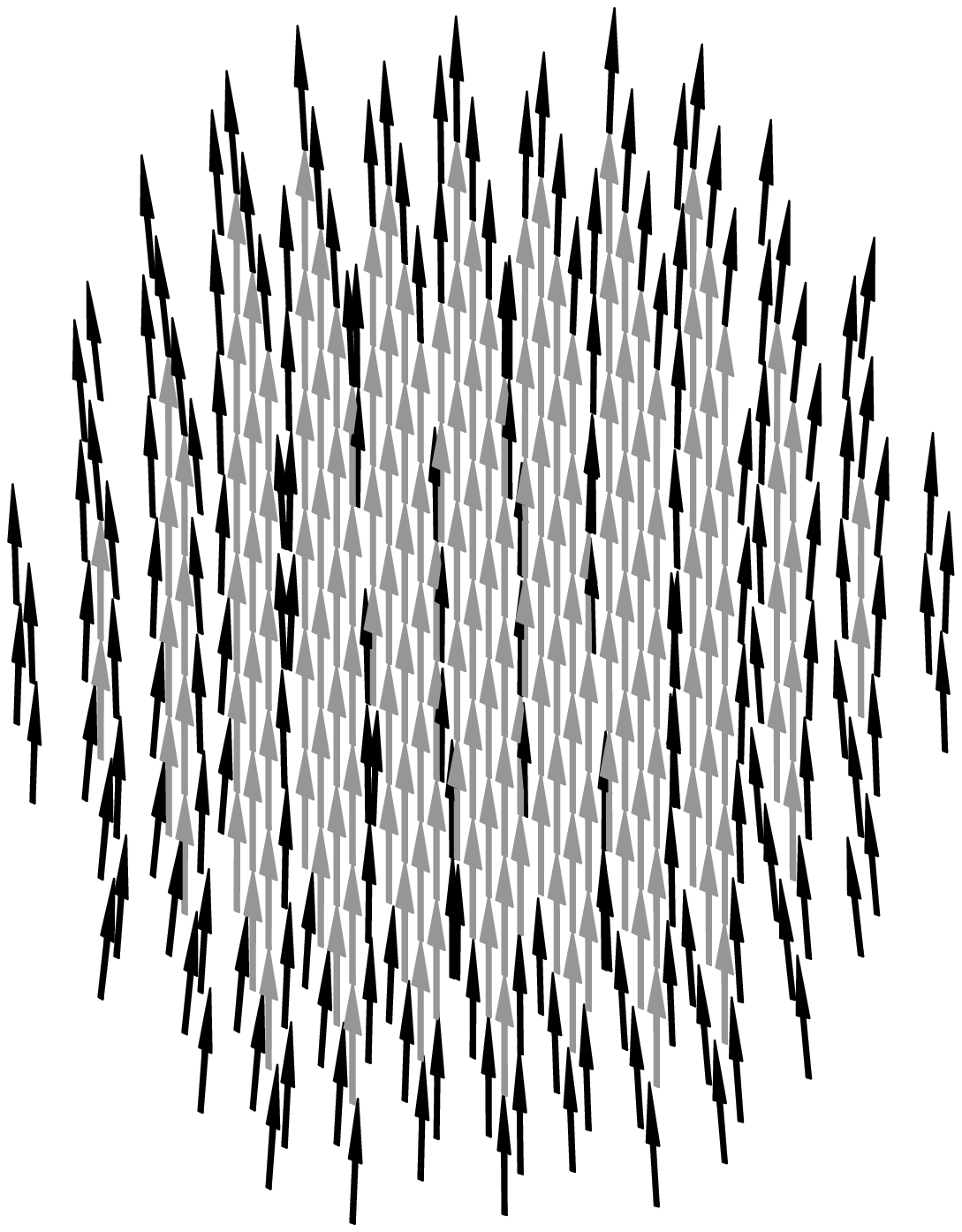, width=4.5cm, height=4.5cm}}
\end{picture}
\vspace{0.5cm}
\caption{\label{structure1}
Magnetic structure for $j=0.1, k_s=1$ for the field values $%
h=-4.0,0,0.64,0.8,0.88,4$ which correspond to the saturation states and
different switching fields shown in Fig.\ \ref{hystk1n10}. These
field values correspond to the pictures when starting from the
upper array and moving right, down left, and then right.
Obviously, grey arrows represent core spins and black arrows represent
surface spins.  
}
\end{figure*}

Now we study the effect of exchange coupling on the hysteresis
loop of a spherical particle containing ${\cal N} = 360$ spins (176
surface spins and 184 core spins). 
We first consider the case in which the anisotropy constants in
the core and on the surface are equal, i.e. $k_{s}=1.0$, and the
magnetic field applied along the easy axis of the core spins, so as 
to investigate the influence of radial direction of surface anisotropy.
For $j\ll 1$, i.e. $j=0,0.01$, we can see along portion 1-2 in Fig.\ \ref{hystk1n10} a
progressive decrease (in absolute value) of the 
magnetization, which is due to the alignment of surface spins, since as
the field direction is along the core easy axis 
the core spins have a rectangular cycle and the jump is at $h=1.0$.
%
%
Next, along portion 2-3 we can see two jumps.
Indeed, according to the distribution of surface easy axes in Fig.\ \ref{number}, and
the critical field as a function of $\psi$ in Fig.\ \ref{hc_width}
(left), those surface spins 
with $\psi_s$ between 0.6 and 1.0 are responsible for the first jump,
and those with $\psi_s$ between 0.4 and 0.6 or 1.0 and 1.2
are responsible for the second jump. Next, along portion 3-4 we have
successive small jumps and thereby a slight decrease of the
magnetization.
The origin of these small jumps resides in two contributions. One
contribution comes from those surface spins whose easy axis makes an angle
around 0.2 with the field. Even though the corresponding height of jump is
large (see Fig.\ \ref{hc_width}, right), their number is rather small
(see Fig.\ \ref{number}) thus rendering a small
contribution to the magnetization. The other contribution is due to surface
spins with an angle $\psi_s\simeq 1.4$, which yield a small
contribution owing to the fact that the height of the corresponding jump is
very small (see Fig.\ \ref{hc_width}, $\psi_s>1.2$), even though their number is relatively
large. On the last portion of the lower branch of the hysteresis in
Fig.\ \ref{hystk1n10}, we see another big jump,
which is due to the switching of core spins at the field $h_{c}=1.0$. At last,
there is a slow increase of magnetization due to a final adjustment of
surface spins along the field direction. In the present case, the
surface fully switches before the core (see Fig.\ \ref{structure1}).\\ 
For $j=0.1$, we see that the surface behavior remains almost the same
as in the previous cases, whereas the core spins now switch
cluster-wise as can be seen in the $4^{th}$ picture of Fig.\
\ref{structure1}. Indeed, regarding the exchange field as a small
perturbation 
of the applied magnetic field,  it is clear that the core spins located near
the surface are subject to an effective field whose direction is slightly deviated
from their easy axis, i.e. the corresponding angle $\psi$ is slightly
different from zero. 
Now, in Fig.\ \ref{hc_width} (left) we can see that this
little deviation in $\psi$ produces an important change in the switching field.
On the contrary, we find that this effect is almost absent in what concerns
the jumping field of surface spins, as can be seen along portion 2-3 in
Fig.\ \ref{hystk1n10} upon comparing the loops for $j=0,0.01$ and
$j=0.1$. Indeed, the surface spins responsible for these jumps have
their easy axes at an angle $0.6<\psi_s<1.0,$ and hence the change in
the corresponding critical field is very small (see Fig.\ \ref{hc_width} left).
In Fig.\ \ref{hystk1n10} we can also see that for $j=0.1$, i.e. when the exchange
energy becomes comparable with anisotropy and Zeeman energy, there are more
jumps that can be attributed to the switching of different spherical shells
of spins starting from the surface down to the center. This situation is
sketched in Fig.\ \ref{structure1}. For example, for $h=0$ one can
see that the exchange has a little influence on surface spins, as they are
directed almost along their easy axes; for $h=0.64$ the surface spins show
the same behavior as in the absence of exchange, but part of core spins,
located near the surface, are deviated from their easy axes. At the
field $h=0.8$ all these core spins have already switched.

\begin{figure*}[floatfix]
\unitlength1cm 
\vspace{0.5cm}
\begin{picture}(4,4.5)(5,-0.5)
\centerline{\psfig{file=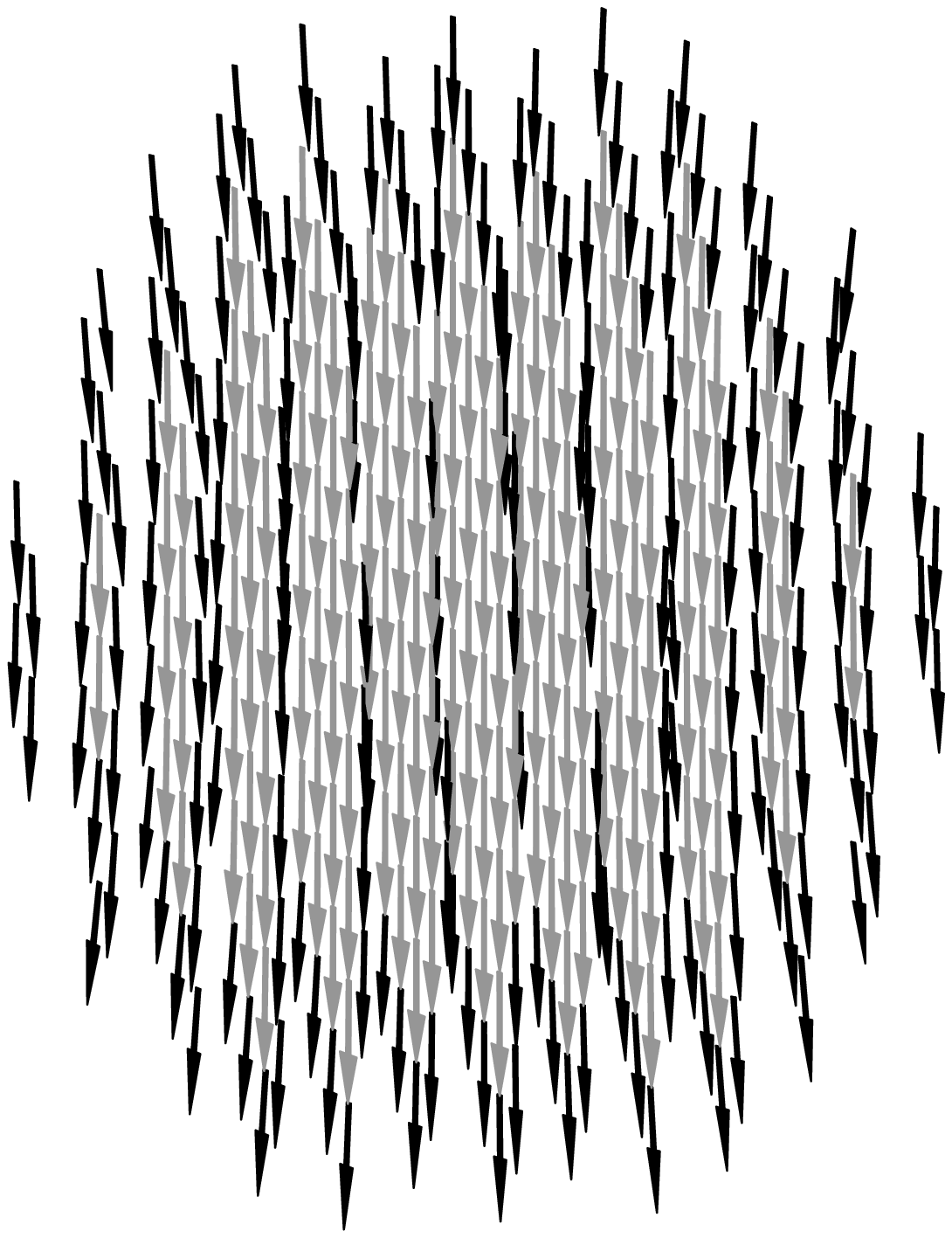, width=4.5cm, height=4.5cm}}
\end{picture}
\begin{picture}(4,4.5)(4,-0.50)
\centerline{\psfig{file=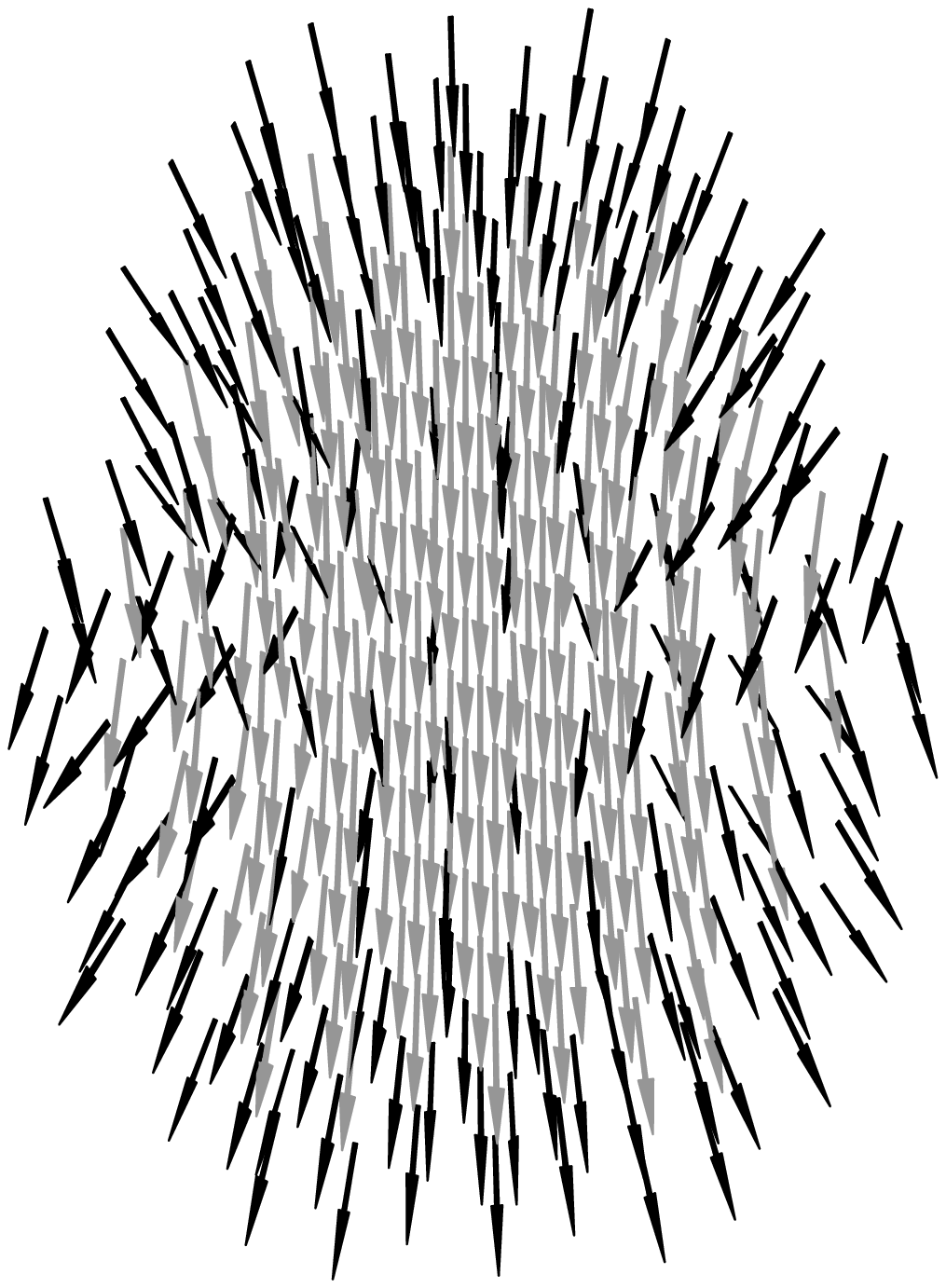, width=4.5cm, height=4.5cm}}
\end{picture}
\begin{picture}(4,4.5)(3.5,-0.5)
\centerline{\psfig{file=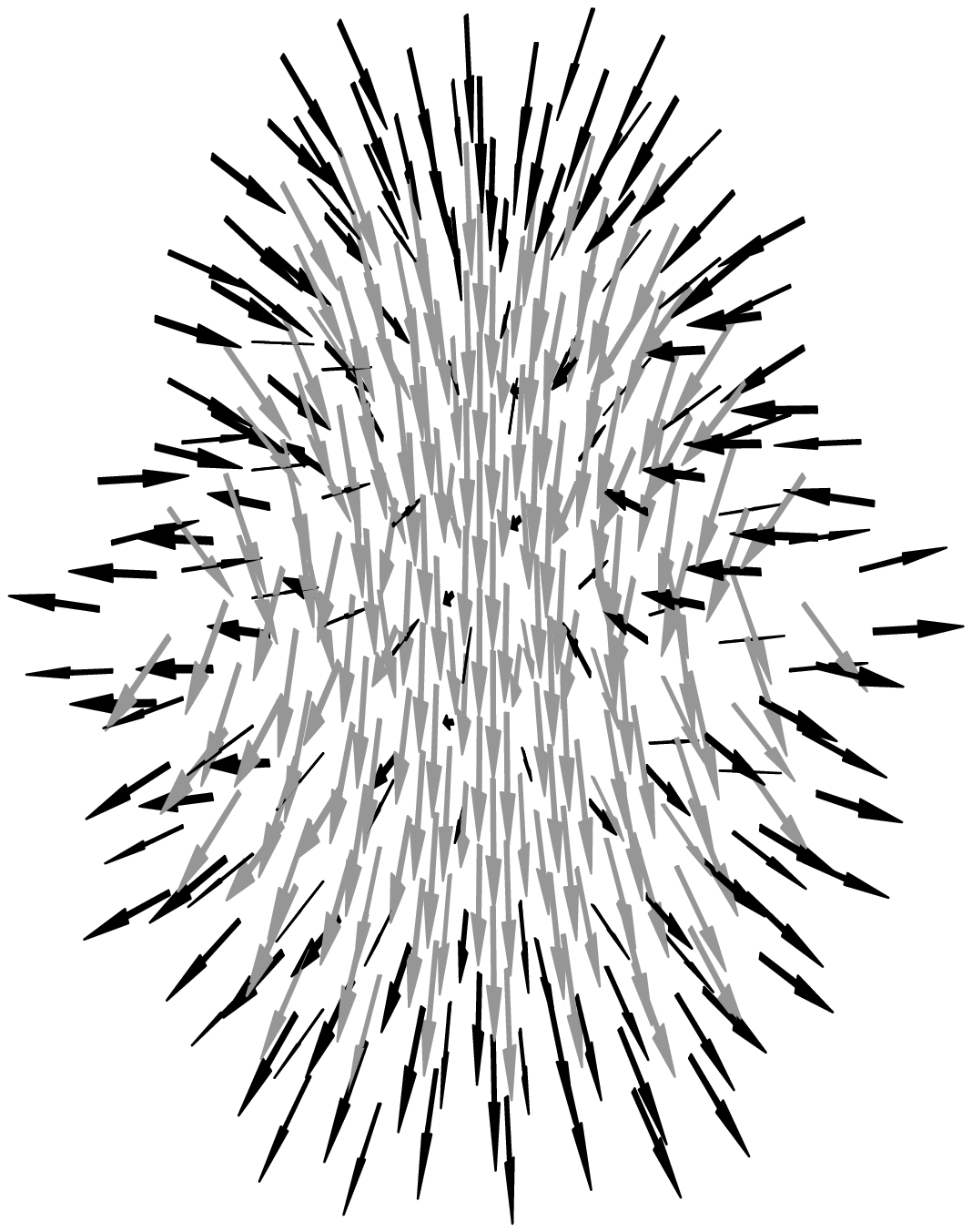, width=4.5cm, height=4.5cm}}
\end{picture}\\
\begin{picture}(4,4.5)(3.5,-0.5)
\centerline{\psfig{file=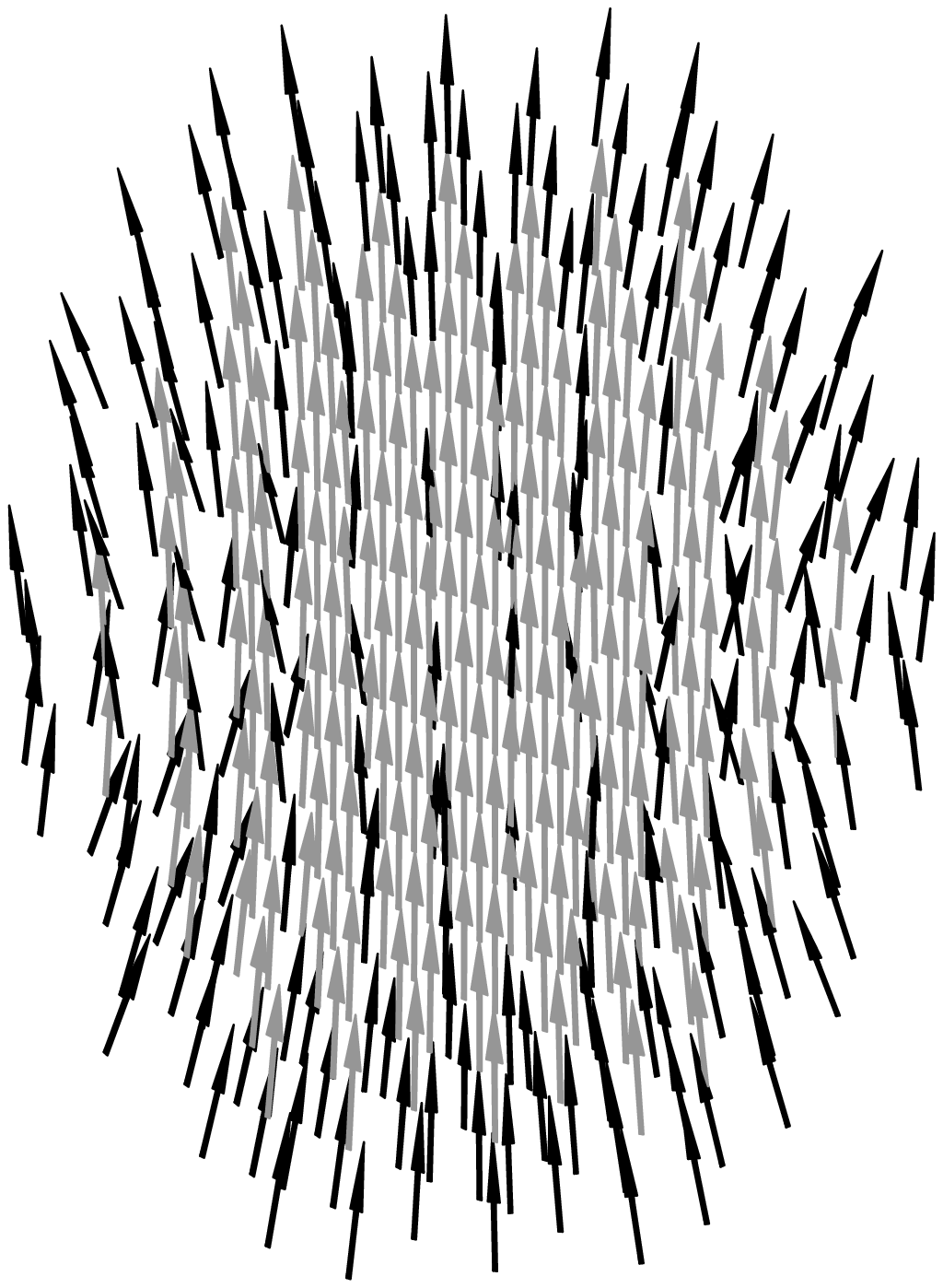, width=4.5cm, height=4.5cm}}
\end{picture}
\begin{picture}(2,4.5)(2.5,-0.5)
\centerline{\psfig{file=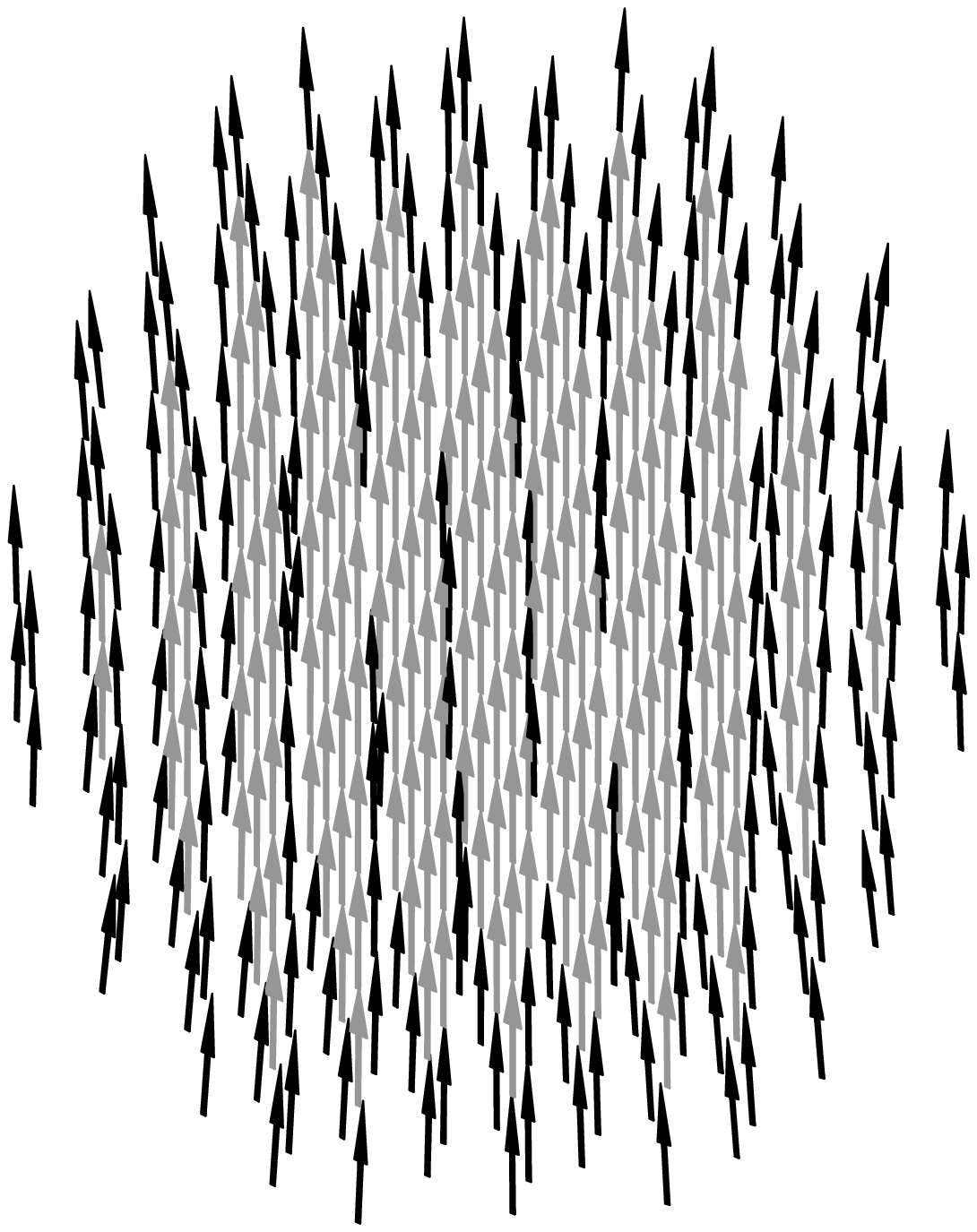, width=4.5cm, height=4.5cm}}
\end{picture}
\caption{\label{structure2}
Magnetic structure for $j=1, k_s=1$ for the field values $%
h=-4.0,0,0.56,0.6,4$ which correspond to the saturation states and different
switching fields shown in Fig.\ \ref{hystk1n10}. As in Fig.\ \ref{structure1},
grey arrows represent core spins and black arrows represent surface spins.
}
\end{figure*}

For $j=1\sim k_s$, even that there is only one jump, the hysteresis
loop is not rectangular owing to the fact that the spins rotate in a non-coherent
way, as can be seen in Fig.\ \ref{structure2}. This is due to a
compromise between anisotropy and exchange energies, see for example
the picture for $h=0$. Moreover, even a small number of neighbors lying
in the core produces a large effect via exchange on the behavior of a
surface spin.
%
%
For much larger values of $j$ the spins are tightly coupled and move
together, and the corresponding (numerically obtained) critical field
$h_{c}$ coincides with 
the (analytical) expression obtained in the limit  $J\longrightarrow
\infty$, i.e. $h_{c}=N_{c}/{\cal N}$, where $N_{c}$ is the number of core spins. 
This expression for $h_{c}$ has been obtained by
summing over the
direction of surface easy axes which results in a constant surface
energy contribution proportional to $k_{s}$. Hence, due to
spherical symmetry, the surface anisotropy constant does not enter
the final expression of $h_{c}$.

\begin{figure*}[!]
\unitlength1cm 
\begin{picture}(7,8)(1,-8)
\centerline{\psfig{file=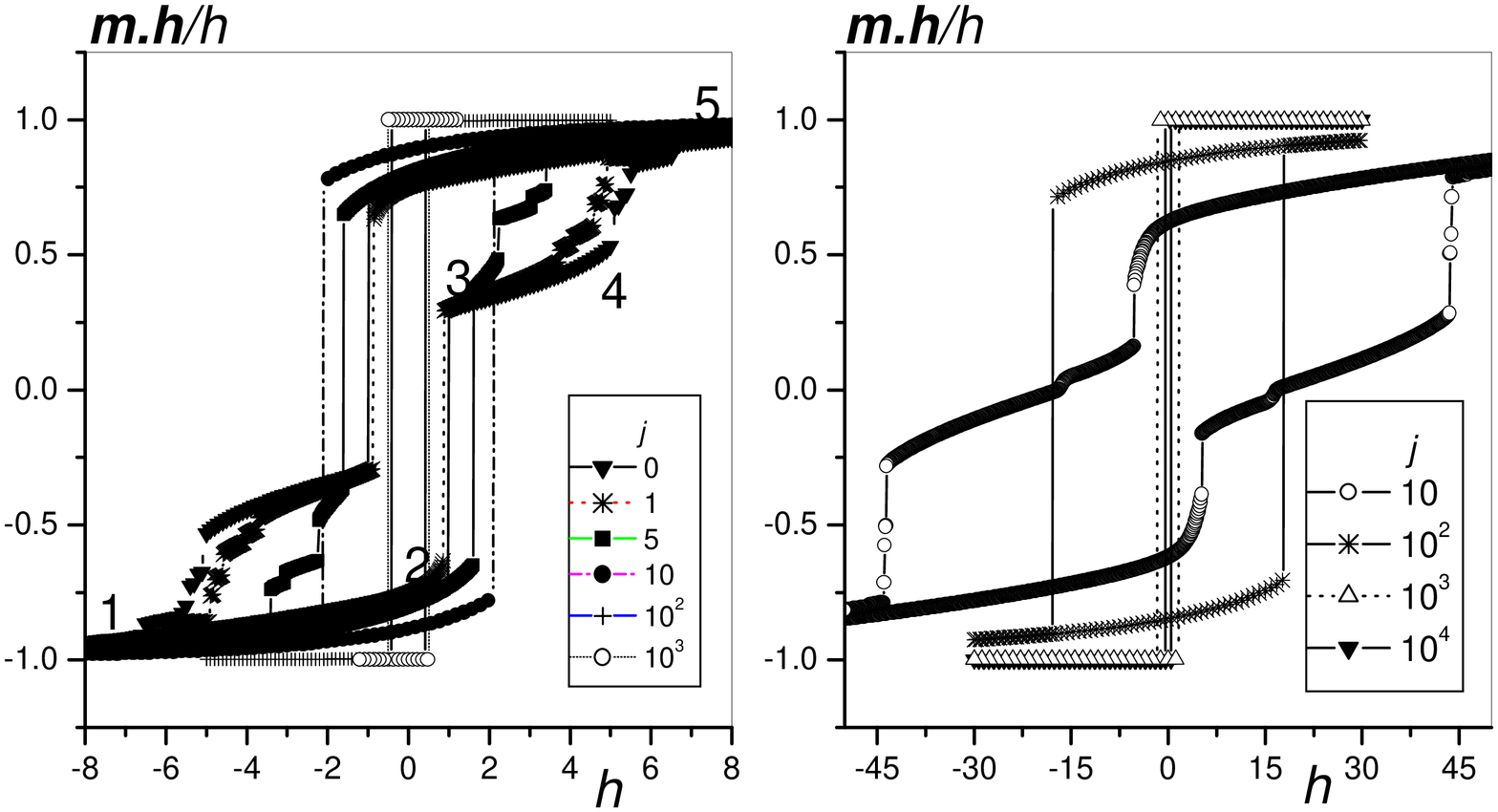,angle=-90,width=14cm}}
\end{picture}
\caption{\label{hystk10n10_k100n7}
Left: Hysteresis loops for $\psi =0$, $k_s=10$ and different values of
$j$. $D=10\,({\cal N}=360)$. Right: Hysteresis loops for $\psi =0$, $k_s=10^2$
and different values of $j$. $D=7\,({\cal N}=123)$.
}
\end{figure*}

Now we consider the case of larger values of $k_s$, e.g. $k_{s}=10$,
so as to investigate the influence of surface
anisotropy both in direction and strength. The results are presented
in Fig.\ \ref{hystk10n10_k100n7} (left).
%
%
Here, a notable difference with respect to the previous case, $k_s=1$,
is the fact that the core now switches before the surface and
at higher fields. Moreover, there
appear more jumps which may be attributed to the switching of various 
clusters of surface spins.
Both cases show that as the ratio $j/k_s$ decreases, the magnetization
requires higher fields to saturate. This is further illustrated by
Fig.\ \ref{hystk10n10_k100n7} (right) where $k_s=10^2=j$ for a smaller particle.

Let us now summarize the ongoing discussion.
We observe that considering a radial distribution for surface anisotropy,
leads, even in the case of very strong exchange, to an important quantitative
deviation from the classical SW model. In particular, the critical field in our model
is given by
\begin{equation}\label{hcvshcsw}
H_{c}^{r}=\frac{N_c}{\cal N}H_{c}^{u},  
\end{equation}
where $H_{c}^{r}$ is the critical field for a spherical particle with radial
anisotropy on the surface and uniaxial in the core, $H_{c}^{u}$ is the critical
field for a spherical particle with uniaxial anisotropy for all spins. 
Therefore, when $j$ and $k_s$ are comparable, the compromise between
exchange coupling, favoring a full alignment of the spins along each other, and
surface anisotropy, which favors the alignment of spins along their
radial easy axes, produces large deviations from the
SW model. More precisely, the shape of the hysteresis loop
is no longer rectangular and there appear multiple jumps. The appearance of
these jumps makes
it necessary to define two field values with the help of which a
hysteresis loop can be characterized. A value that
marks the limit of metastability, called the {\it critical field}, and the
other value which marks the magnetization reversal, i.e. when the
projection of the magnetization on the field direction changes sign,
and this is why it is called the {\it switching field} (or still coercive field). 

\subsection{Effect of the particle's size $\cal N$}
\begin{figure}[h]
\unitlength1cm 
\begin{picture}(8,10)(-1.5,-10.5)
\centerline{\psfig{file=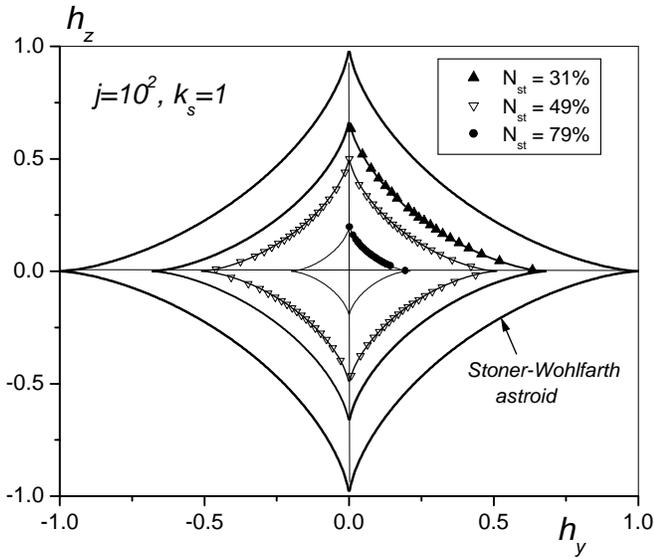,angle=-90,width=12cm}}
\end{picture}
\vspace{-2.5cm}
\caption{\label{ast_size}
Astroid for $k_{s}=1, j=10^2$ for different values of the
surface-to-volume ratio $N_{st}\equiv N_s/{\cal N}$.
The lines on the astroids inside the SW one are only guides for
the numerical data.
}
\end{figure}

Here, we study the effect of varying the particle's size while keeping
$j$ and $k_s$ fixed. So we use the same value of anisotropy constant
for all spins and strong exchange, i.e. $k_s = 1, j = 10^2$, and vary the
particle's diameter from 6 (${\cal N} = 56$) to 30 (${\cal N}
= 12712$). 
%
\begin{figure*}[!]
\unitlength1cm 
\begin{picture}(7,8)(1,-8)
\centerline{\psfig{file=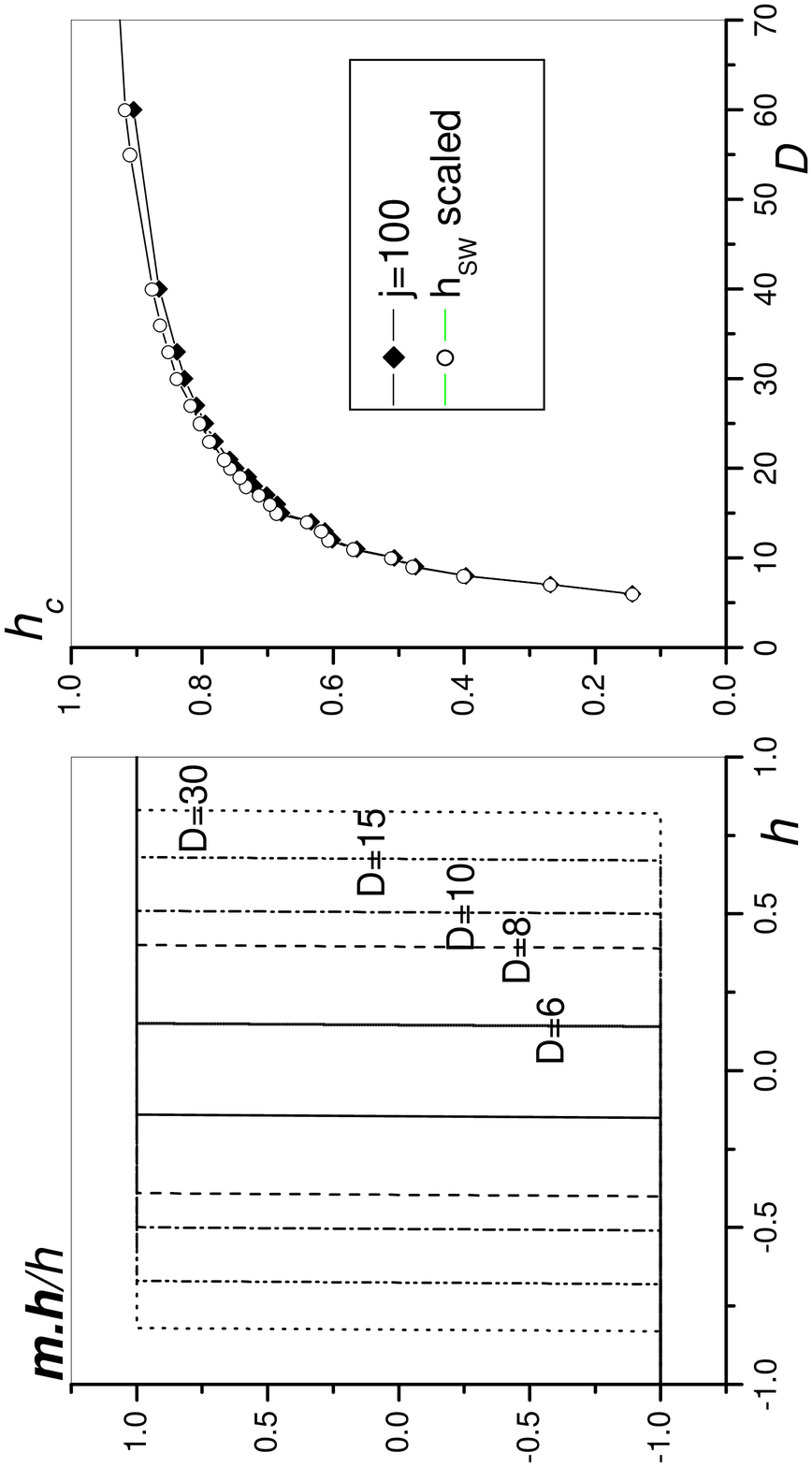,angle=-90,width=14cm}}
\end{picture}
\caption{\label{hyst_hc_size}
Left: Hysteresis loops for $\psi = 0, k_s=1, j=10^2$
for different values of the particle's diameter $D$.
Right: (in diamonds) Switching field for the same parameters as a function of
$D$. $h_{SW}(N_c/{\cal N})$ is the SW switching field
multiplied by the relative number of core spins.
}
\end{figure*}
%
In Fig.\ \ref{hyst_hc_size} (left) are presented hysteresis cycles of a particle
with different diameters 
when the field is along the core easy axis, and on the right the variation
with the particle's 
diameter of the critical field\cite{footnote1} (in diamonds) obtained from
the numerical solution of the 
Landau-Lifshitz equation for $j=10^2$,  and (in circles) the SW critical field
multiplied by the core-to-volume ratio (see Eq. (\ref{hcvshcsw})).
The figure on the left shows that for such a value of $k_s$ the
hysteresis loop is rectangular for all sizes, and that the critical field 
decreases with the particle's size. The latter fact is clearly
illustrated by the plot on the right, which also shows that for
$k_s=1=10^{-2}j$, all these hysteresis loops can be scaled with those
rendered by the SW model. 
Next, Fig.\ \ref{ast_size} shows the variation with the surface-to-volume
ratio $N_{st}\equiv N_s/{\cal N}$ 
of the critical field for all angles between the core easy axis and
magnetic field, this is the limit-of-metastability curve. 
These results show that, even in the general case of a field applied at
an arbitrary angle with respect to the core easy axis, the critical
field of a spherical particle with $k_s=1$ can be obtained from the
SW model through a scaling with constant $N_c/{\cal N}$.
One should also note that the astroid for all
particle sizes falls inside that of SW, in accordance
with Fig.\ \ref{hyst_hc_size} (right), and the larger the surface contribution
the more the astroid shrinks.

Therefore, for $k_s=1$ our results for the hysteresis loop and
limit-of-metastability curve can be
scaled with those of SW model with the scaling constant
$N_c/{\cal N}$, which is smaller than $1$ for a particle of any finite
size.
%
%

\begin{figure*}[ht]
\unitlength1cm 
\begin{picture}(10,10)(0,-10)
\centerline{\psfig{file=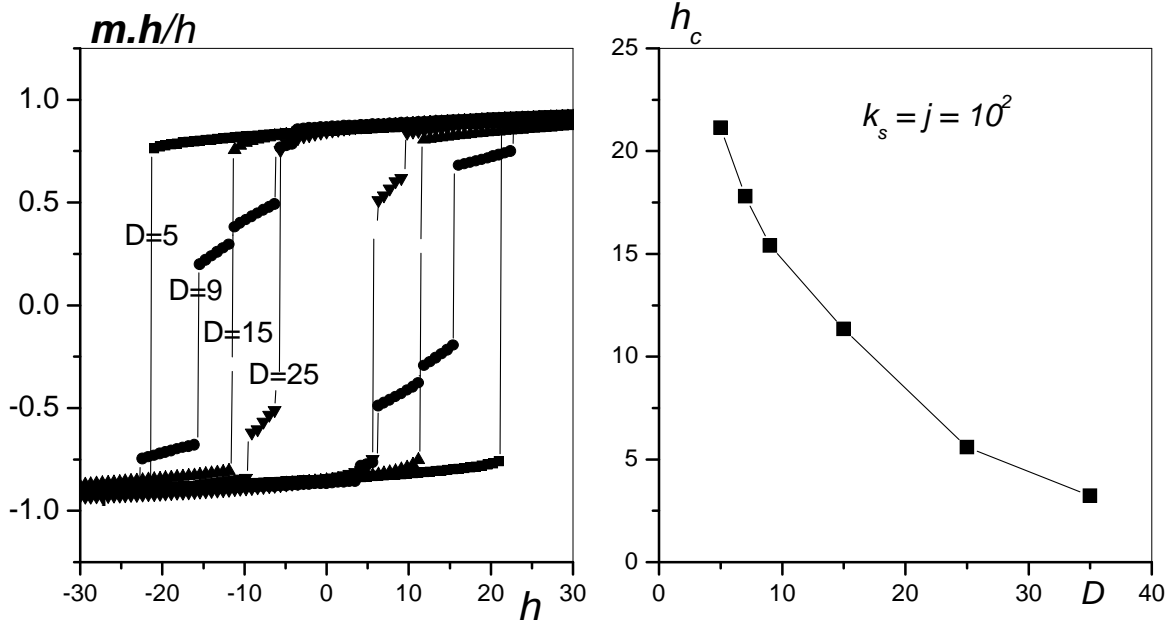,angle=-90,width=16cm}}
\end{picture}
%
\caption{\label{hchyst_j100ks100}
Left: Hysteresis cycle for $\psi=0, j=k_s=10^2$, and 
different values of the particle's diameter $D$. Right: Switching
field as a function of $N$ for the same parameters. 
}
\end{figure*}

Next, in Fig.\ \ref{hchyst_j100ks100} (left) we present the hysteresis loop in the
case where the surface anisotropy constant $k_s$ equals
the exchange coupling and the field is applied along the core
easy axis, and in Fig.\ \ref{hchyst_j100ks100} (right) the
switching field\cite{footnote2} as a function of
the particle's diameter $D$. There are two new features in
comparison with the previous case of $k_s=1$: the values of the
switching field are much higher, and more importantly, its behavior
as a function of the particle's size
is opposite to that of the previous case. Indeed, here we see that this field
increases when the particle's size is lowered. For such high values of
$k_s$ ($K_s \gg K_c$) surface spins are aligned along their easy axes, and because of
strong exchange coupling they also drive core spins in their switching
process. Thus, the smaller is the particle the larger is the surface contribution,
and the larger is the field required for complete reversal of the
particle's magnetization. This could explain the non-saturation of 
magnetization that has been observed in e.g. cobalt particles\cite{Respaud}. 
%
%
\subsection{Effect of the surface anisotropy constant $k_s$}
Now, we fix the exchange coupling constant $j$, the particle's total number
of spins ${\cal N}$, and
vary the surface anisotropy constant $k_s$. Because $K_{c}$ is in
general 2 to 3 orders of magnitude smaller than $J$, we have investigated the
effect of surface anisotropy constant in the case of $j=J/K_c=10^{2}$. 

\begin{figure}[ht!]
\unitlength1cm 
\begin{picture}(9,9)(-1.5,-9)
\centerline{\psfig{file=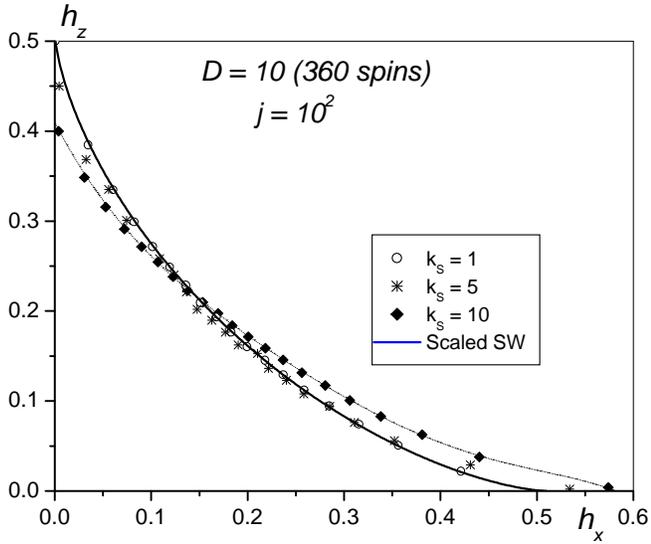, angle=-90, width=12cm}}
\end{picture}
\vspace{-1cm}
\caption{
Astroid for $j=10^2$, ${\cal N}=360$ and different values of surface
anisotropy constant $k_{s}$. The full dark line is the SW
astroid scaled with $N_c/{\cal N}$, but the dotted line is only a
guide for the eye.
}\label{ast_ks}
\end{figure}

In contrast with the case $k_{s}=1$ and $j=10^2-10^3$ where the 
hysteresis loop and the limit-of-metastability curve scale with the
SW ones with the same scaling constant for all angles between the
applied field and core easy axis, we find that for $1<k_s<20$ the scaling constant
depends on the angle $\psi$, as can be seen in Fig.\ \ref{ast_ks}. 
This fact explains the deformation of the SW astroid, that is a
depression in the core easy direction and an enhancement in the
perpendicular direction. 

\begin{figure*}[t!]
\unitlength1cm 
\begin{picture}(12,12)(0,-10)
\centerline{\psfig{file=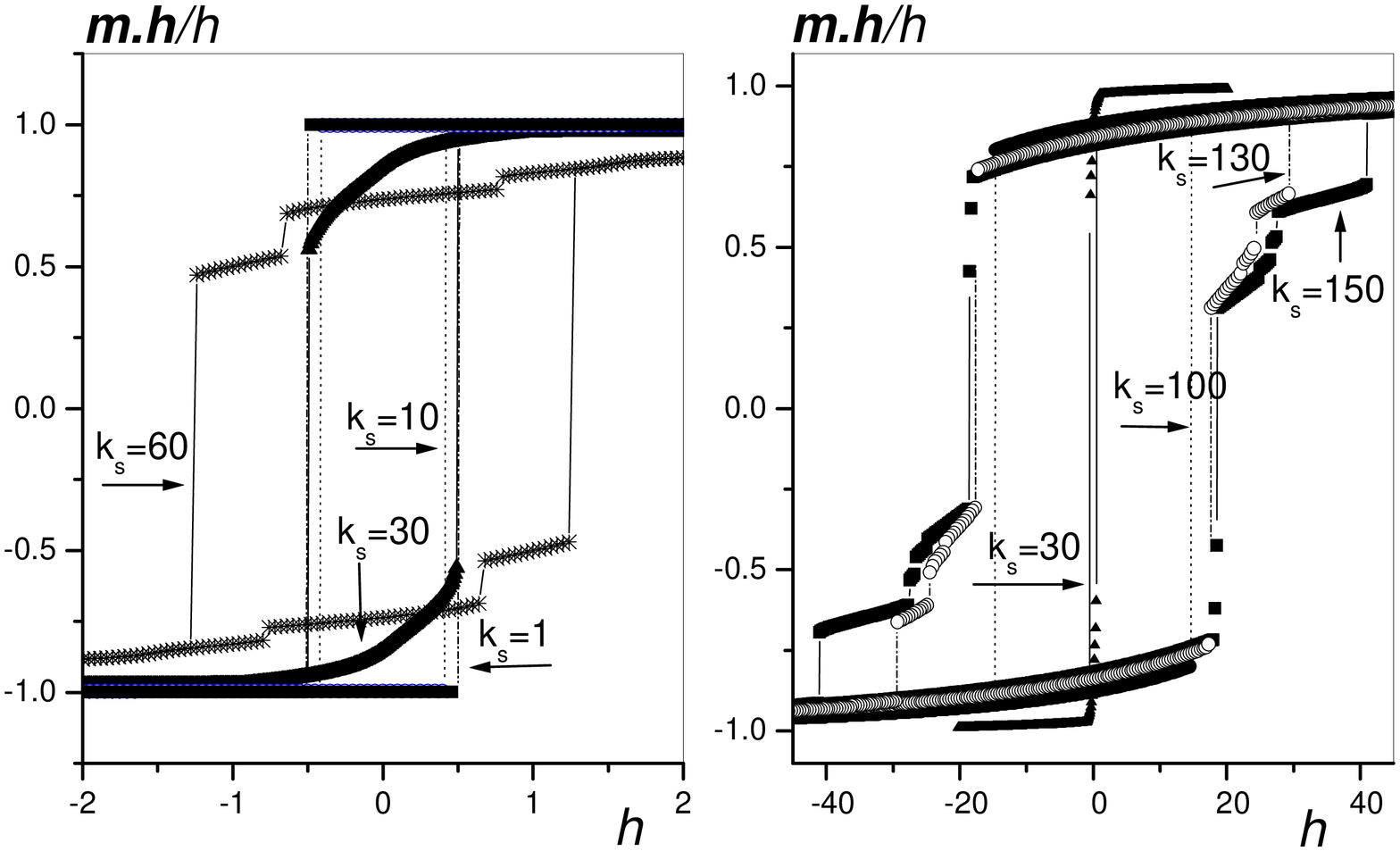,angle=-90,width=15cm}}
\end{picture}
%
\caption{\label{hysts_ks_a0}
Hysteresis loop for $\psi =0$, $j=10^2$, $D=10$ and
different values of surface anisotropy constant $k_{s}$.
These two sets of data can not be presented as one plot because of scaling
mismatch.
}
\end{figure*}

\begin{figure}[h]
\unitlength1cm
\begin{picture}(10,10)(-1,-9)
\centerline{\psfig{file=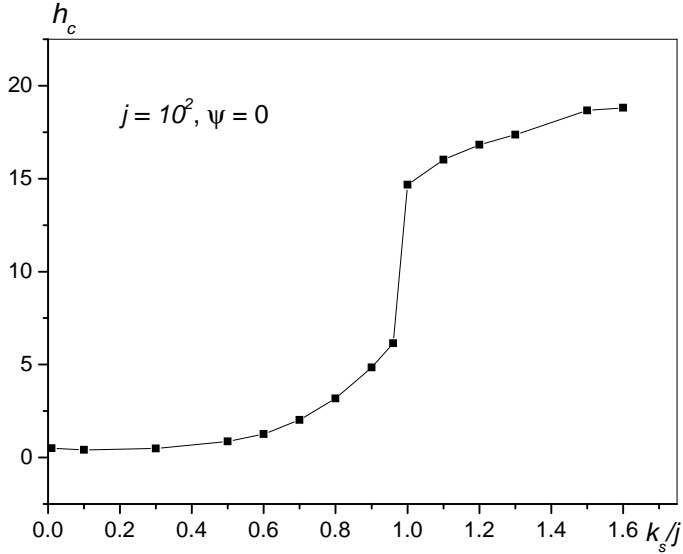, angle=-90, width=12cm}}
\end{picture}
%
\caption{\label{hcvsks} 
Switching field versus the surface anisotropy constant for $\psi =0$,
$j=10^2$, and $D=10$.
}
\end{figure}

For larger values of $k_{s}$ we have computed the hysteresis loop for
$\psi=0, {\cal N}=360, j=10^2$. The results are given in Fig.\
\ref{hysts_ks_a0}.
%
%
Here, we first note that the shape of the hysteresis loop is rather different from
that rendered by the SW model, since for $k_s=30$, for instance, the
hysteresis loop is no longer rectangular, even that $\psi=0$. As
explained earlier, this effect is due to the now more pronounced non-uniform rotation of
surface spins and core spins located near the surface, and thereby
that of the particle's magnetization. This non-uniform switching
process causes large deviations from the SW model, and thereby no scaling
with the latter is possible.
%
%
From Fig.\ \ref{hysts_ks_a0}, we extract and plot in Fig.\
\ref{hcvsks} the switching field $h_c$ as a function of $k_s/j$,
denoted by $\tilde{k}_s$ in the sequel. We find that 
$h_c$ first slightly decreases for $\tilde{k}_s \lesssim 0.1$ and then increases, and 
when $\tilde{k}_s$ approaches $1$ it jumps to large
values. As discussed above, for such high values of $k_s$ surface spins are
aligned along their easy axes, and because of strong exchange coupling
they also drive core spins in their switching process, which then
requires a very strong field to be completed. 
Clearly, this particular value of $\tilde{k}_s$, to be denoted by
$\tilde{k}_s^c$ ($=1$, here)
marks the passage from a regime where scaling with 
the SW results is possible (either with a $\psi$-dependent or independent
constant) to the second regime where this scaling is no longer
possible because of completely different switching processes. 

Now we present additional data which show that the ``critical value'' ${\tilde k}_s^c$
introduced above depends on (at least) two parameters. 
These are the surface-to-core ratio of exchange coupling $j_s/j$ and the angle $\psi$ at 
which the field is applied with respect to the core easy axis.   

\begin{figure}[h]
\unitlength1cm
\begin{picture}(9,10)(-1,-10)
\centerline{\psfig{file=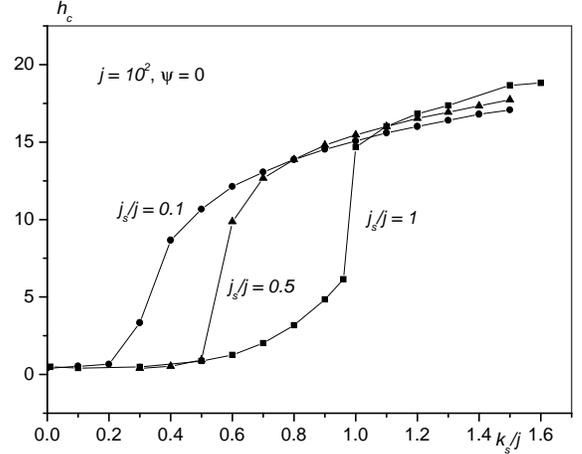, angle=-90, width=10cm}}
\end{picture}
\vspace{-3cm}
\caption{\label{hcvsks2} 
Switching field versus the surface anisotropy constant for $\psi =0$,
and different values of surface-to-core ratio of exchange couplings; $D=10$. 
}
\end{figure}

Let us first discuss the effect of (intra) surface exchange coupling. 
In real materials such as maghemite, it was argued in \cite{Kachkachi} on account of 
M\"ossbauer spectroscopy that $j_s/j<1$.
In Fig.\ \ref{hcvsks2} we have plotted the results for $h_c$ 
obtained with surface exchange coupling $j_s\equiv J_s/K_c$ smaller 
than $j$, i.e. core-core and core-surface couplings. 
First, we see that the "critical" value $\tilde{k}_s^c$ of $\tilde{k}_s$ separating the two
regimes discussed above, decreases with the ratio $j_s/j$. 
This is a consequence of the fact that when $j_s/j < 1$,
surface spins align more easily along their (radial) anisotropy axes
since now they experience a weaker effective field.
We also note that the jump becomes smoother.
Next, if we consider the curve $j_s/j=1$ together with any other curve with
$k_s/j<1$, we see that when $\tilde{k}_s < 1$ the switching field
is larger for $j_s < j$ than for $j_s=j$, and the opposite holds when
$\tilde{k}_s > 1$.

To understand this result, let us imagine a particle containing (at
least) two groups of surface spins, a group 1 with exchange coupling
$j_s=j$ and group 2 with $j_s<j$.
When $\tilde{k}_s^2 < \tilde{k}_s < \tilde{k}_s^1$, $\tilde{k}_s^i$
being the critical value of $\tilde{k}_s$ for group $i$, the spins in
group 1 are of SW type, while those of group 2 are of non-SW type,
in the sense that they switch in a coherent way or cluster-wise,
respectively. 
Hence, as demonstrated earlier, the reversal of spins in group 2 always 
requires a larger switching field.
%
%
On the other hand, when $k_s$ exceeds the largest exchange coupling in
the particle, i.e. $j$, the switching field of the whole particle
decreases with $j_s/j$. Now the spins of both groups are of
non-SW type, and their switching operates cluster-wise, but obviously
the latter requires a higher applied field for group 1 than for group 2.
%

\begin{figure}[h!]
\unitlength1cm 
\begin{picture}(10,9)(-1,-9.5)
\centerline{\psfig{file=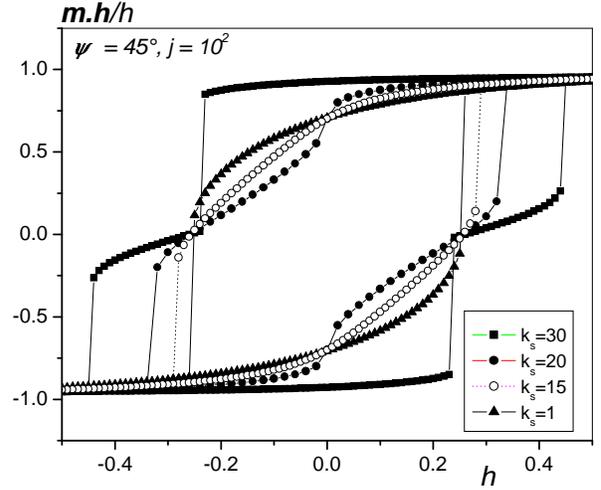,angle=-90, width=11cm}}
\end{picture}
\vspace{-2cm}
\caption{Hysteresis for $\psi =45{{}^{\circ }}, j=10^2, D=10 ({\cal N}=360)$ and
different values of surface anisotropy constant $k_{s}$.
}
\label{hyst_ks_a45}
\end{figure}

Next, a similar effect is obtained when the field is applied at an
arbitrary 
angle with respect to the core easy axis, as is the case for instance in an assembly of 
nanoparticles. Here, we consider the case of $\psi=\pi/4$.
We find that there appear multiple large jumps at a smaller value of ${\tilde k}_{s}$ 
($\sim 0.2$), as can be seen in Fig.\ \ref{hyst_ks_a45}.

For an order of magnitude estimate of $K_s$ and the critical (or
saturation) field, consider a 4 nm cobalt
particle of fcc crystal structure, for which the lattice spacing is
$a=3.554\,\AA$, 
and there are 4 cobalt atoms per unit cell. The (bulk) magneto-crystalline
anisotropy is $K_c\simeq 3\times 10^{-17}$ erg/spin or $2.7\times 10^6$
erg/cm$^3$, and the saturation magnetization is $M_s\simeq 1422$
emu/cm$^3$.
The critical field is given by $H_c=(2K_c/M_s)h_c$.
For $\psi=0$, $\tilde{k}^c_s=1$ and $h_c=15$, so $H_c\simeq 6\,$T. 
On the other hand, $\tilde{k}^c_s=1$ means that the effective exchange field
experienced by a spin on the surface is of the order of the anisotropy
field, i.e. $zSJ/2\sim 2K_s$. Then using $J\simeq8\,$mev we get
$K_s\simeq5.22\times 10^{-14}\,$erg/spin, or using the area per surface spin
(approximately $a^2/8$), $K_s\simeq 5\,$erg/cm$^2$.
For the case of $\psi=\pi/4$, $\tilde{k}^c_s\simeq 0.2$ and $h_c\simeq 0.3$,
which leads to $H_c\simeq 0.1\,$T and $K_s\simeq 1.2\times
10^{-14}\,$erg/spin or $1.2\,$erg/cm$^2$. 
  
%
%
\section{Conclusion}
Our model of a spherical particle with uniaxial
anisotropy in the core and radial anisotropy on the surface leads to
mainly two pertinent regions for the surface anisotropy constant
$k_s$, with $k_s > 1$ ($K_s>K_c$): 

\begin{itemize}
\item
For small values of this parameter, e.g. $k_s/j\sim 0.01$ our model
renders hysteresis loops and limit-of-metastability curves that scale with the
SW results for all values of the angle $\psi$
between the core easy axis and the applied field, the scaling constant
being $N_c/{\cal N}$, which is smaller than $1$ for a particle of any finite
size. On the other hand, the critical field, which coincides in the
present case with the switching field, increases with the
particle's size and tends to the SW critical field in very large
systems, and thereby the corresponding astroid falls inside the SW astroid
for all particle sizes.

For larger values of $k_s/j$, 
but $k_s/j \lesssim 0.2$, we still have the same kind of scaling but
the corresponding constant depends 
on $\psi$. This is reflected by a deformation of the
limit-of-metastability curve. More 
precisely, the latter is depressed in the core easy
direction and enhanced in the perpendicular direction. However, there
is still only one jump in the hysteresis loop implying that the
magnetization reversal can be considered as uniform.

\item
For much larger values of $k_s/j$, starting from $k_s/j\simeq 1$,
there appear multiple steps in the hysteresis loop
which may be associated with the switching of spin clusters. The
appearance of these steps makes the calculated hysteresis loops both
qualitatively and quantitatively 
different from those of SW model, as the magnetization
reversal can no longer be considered as uniform, and one has then to define two
characteristic values of the field associated with a hysteresis loop:
the {\it critical field} and {\it the switching field}.
In addition, in the present case, there are two more new features: the values of the 
switching field are much higher than in SW model, and more importantly, its behavior
as a function of the particle's size is opposite to that of the
previous cases. More precisely, here we find that this field 
increases when the particle's size is lowered. This is in agreement with
the experimental observations in nanoparticles (see e.g. \cite{Chenetal}
for cobalt particles).
\end{itemize}

Therefore, assuming radial anisotropy on the surface, we find that
there is a ``critical'' value $(K_s/J)^c$ of the ratio 
$K_s/J$ beyond which, large deviations are observed with 
respect to the SW model in the hysteresis loop and thereby
the limit-of-metastability curve, since in this case the magnetization
reverses its direction in a non-uniform manner via a progressive switching of
spin clusters. 
So, in order to deal with these new features one
has to resort to microscopic approaches such as the one used in
this work.
In fact, it is found that the critical value $(K_s/J)^c$ is even
smaller for smaller surface-to-core ratios of exchange coupling and larger
angles between the applied magnetic field and the core easy direction, as
it is more likely in realistic materials.   

In a subsequent work we apply the present method to
cubo-octahedral cobalt particles with a diameter of ca. 3 nm recently
studied in \cite{Jamet1} (see also \cite{nanomaterials} for Pt particles). 
These are particles with fcc structure and truncated octahedrons on the surface, in which the core
has a cubic anisotropy, and the surface
anisotropy easy axes are believed to be along edges and facets with different
constants $K_{s}^{\alpha }$ but whose values are uncertain at
present. 
In our calculations we vary these parameters and study the effect of
surface anisotropy on the Stoner-Wohlfarth astroid that has been
experimentally measured in \cite{Jamet1}, where these anisotropy
constants have been estimated from magnetic measurements. The final outcome
of our calculations should give an estimation of $K_{s}^{\alpha }$ by
comparing with these experimental results. 
Another related issue of particular
interest to us is the fact that these fcc particles
(see \cite{Jamet1} for cobalt and \cite{Kitakami} for iron) seem to
exhibit an effective {\it uniaxial} anisotropy despite their cubic
crystal symmetry. This work is in progress.  

\section*{Acknowledgments}
We thank D.A. Garanin and M. Nogu\`es for reading the manuscript and suggesting
improvements. M. Dimian thanks the Laboratoire de Magn\'etisme et d'Optique for the hospitality extended to him
during his training under the Socrates program, 1 March - 31 July 2001.


\begin{thebibliography}{15}
\bibitem{footnote3} It is sufficient to consider
such a small particle, since for a strong exchange coupling and all easy
axes along the same direction $z$ the result would be the same for a large
particle, but then the calculation of the SW astroid is much more time
consuming.
\bibitem{footnote1} For $k_s=1$, there is only one jump in
the hysteresis loop as can be seen in Fig.\ \ref{hyst_hc_size} (left). Hence, the
critical field coincides with the switching field.
\bibitem{footnote2} This is different from the critical field in
the present case of $k_s=10^2$.
\end{thebibliography}
\end{document}